# Coupled states of electromagnetic fields with magnetic-dipolar-mode vortices: MDM-vortex polaritons

E.O. Kamenetskii, R. Joffe, and R. Shavit

Department of Electrical and Computer Engineering,
Ben Gurion University of the Negev, Beer Sheva, Israel

December 6, 2010

**Abstract**

Under the influence of the material environment, electromagnetic fields in the near-field regime exhibit quite different nature from those in the far-field free space. A coupled state of an electromagnetic field with an electric or magnetic dipole-carrying excitation is well known as a polariton. Such a state is the result of the mixing of a photon with an excitation of a material. The most discussed types of polaritons are phonon-polaritons, exciton-polaritons, and surface plasmon-polaritons. Recently, it was shown that in microwaves strong magnon-photon coupling can be achieved due to magnetic-dipolar-mode (MDM) vortices in small thin-film ferrite disks. These coupled states can be specified as MDM-vortex polaritons. In this paper we study properties of MDM-vortex polaritons. We show that MDM-vortex polaritons are characterized by helicity behaviors. For the observed frequency splits of MDM resonances there are different-type helicities. In the split-resonance states one has or localization, or cloaking of electromagnetic fields. We analyze numerically a variety of the field topological structures of MDM-vortex polaritons and give theoretical insights into the possible origin of such topologically distinctive states. The shown properties of MDM-vortex polaritons can be useful for realization of novel microwave metamaterial structures and near-field sensing applications.

**I. Introduction**

The coupling between photons and magnons in a ferromagnet has been studied in many works over a long period of time. In an assumption that there exists an oscillating photon field associated with the spin fluctuations in a ferromagnet, one can observe the photon-like and magnon-like parts in the dispersion relations. The dispersion characteristics for the coupled magnon-photon modes were analyzed for various directions of the incident electromagnetic wave vector and it was found, in particular, that there are the reflectivity bands for electromagnetic radiation incident on the ferromagnet-air interface [1 – 4]. As one of the most attractive effects in studies of the reflection of electromagnetic waves from magnetic materials, there is observation of a nonreciprocal phase behavior [5].

In a general case of oblique incidence on a single ferrite/dielectric interface, apparently different situations arise by changing the directions of incident waves and bias, and incident side of the interface. The solutions obtained for different electromagnetic problems of ferrite-dielectric structures show the time-reversal symmetry breaking (TRSB) effect [6 – 10]. Microwave resonators with the TRSB effect give an example of a nonintegrable electromagnetic system. In general, the concept of nonintegrable, i.e. path-dependent, phase factors is considered as one of the fundamental aspects of electromagnetism. The path-dependent phase factors are the reason of appearance of complex electromagnetic-field eigenfunctions in resonant structures with enclosed ferrite samples, even in the absent of dissipative losses. In such structures, the fields of eigen oscillations are not the fields of standing waves in spite of the fact that the eigen frequencies of a cavity with a ferrite sample are real [11]. Because of the TRSB effect and the complex-wave behaviors, one can observe induced electromagnetic vortices in microwave resonators with ferrite inclusions [12 – 14].

Very interesting effects appear when an oscillating photon field is coupled with a resonant collective-mode behavior of spin fluctuations in a confined ferromagnetic structure. This concerns, in particular, a microwave effect of strong coupling between electromagnetic fields and long-range magnetic dipolar oscillations. Such oscillations, known as magnetic-dipolar-mode (MDM) or magnetostatic (MS) oscillations, take place due to long-range phase coherence of precessing magnetic dipoles in ferrite samples. The wavelength of MDM oscillations is two-four orders of magnitude less than the free-space electromagnetic wavelength at the same microwave frequency [11]. The fields associated with MDM oscillations in confined magnetic structures decay exponentially in strength with increasing distance from the ferrite-vacuum interface. In general, these modes are nonradiative. The nonradiative character of MDMs has two important consequences: (i) MDMs cannot couple directly to photon-like modes (in comparison with photon-like modes, the MDM wavevectors are too great), and (ii) the fields associated with MDMs may be considerably enhanced in strength in comparison with those used to generate them. The electromagnetic radiation only emerges after it has multiply "bounced round" in the structure, during which some energy is lost by absorption to the ferrite material. In a region of a ferromagnetic resonance, the spectra of MDMs strongly depend on geometry of a ferrite body. The most pronounced resonance characteristics one can observe in a quasi-2D ferrite disk. The coupling between an electromagnetic field in a microwave cavity and MDM oscillations in a quasi-2D ferrite disk shows a regular multiresonance spectrum of a high-quality factor [15, 16]. Recently, it was shown that small ferrite disks with MDM spectra behave as strong attractors for electromagnetic waves at resonance frequencies of MDM oscillations [17]. It was found that the regions of strong subwavelength localization of electromagnetic fields (subwavelength energy hot-spots) appear because of unique topological properties of MDM oscillations – the power-flow eigen vortices [17 – 19]. Because of the MDM vortices, one has strong magnon-photon coupling in microwaves. Such coupled states can be specified as the MDM-vortex polaritons.

Unique topological properties of MDMs are originated from nonreciprocal phase behaviors on a lateral surface of a ferrite disk. A numerical analysis of classical complex-wave fields in a ferrite disk gives evidence for the Poynting-vector vortices and the field rotation inside a ferrite disk at frequencies corresponding to the MDM resonances [17 – 19]. The rotation angle of the polarization plane of electromagnetic fields, evident from numerical studies, is represented by a geometrical phase. Manifestation of a geometrical phase in wave dynamics of confined classical structures is well known. For example, the Berry phase for light appears in a twisted optical fiber in which the trajectory of the wave vector makes a closed loop. In this case, the polarization plane rotates during propagation, and the rotation angle is represented by a Berry phase [20]. Due to a Berry phase one can observe a spin-orbit interaction in optics. In particular, it was shown that a spin-orbit interaction of photons results in fine splitting of levels in a ring dielectric resonator, similarly to that of electron levels in an atom [21]. In our case, the geometrical phase of electromagnetic fields appear due to space- and time-variant subwavelength (with respect to free-space electromagnetic fields) magnetization profiles of MDMs in a ferrite disk [22].

The purpose of this paper is to study scattering of microwave electromagnetic fields from MDM-vortex polaritons. A geometrical phase plays a fundamental role in forming the coupled states of electromagnetic fields with MDM vortices. Because of the intrinsic symmetry breakings of the vortex characteristics, a small ferrite particle with a MDM spectrum behaves as a point singular region for electromagnetic waves. Based on a numerical analysis of classical complex-wave fields, we show that due to the "spin-orbit interaction" MDM resonances have frequency splits. For the split states, one has or localization, or cloaking of electromagnetic fields. Definite-phase relationship between the incident electromagnetic wave and microwave magnetization in the MDM particle results in asymmetry in the forward and backward scattering of electromagnetic waves. The broken reflection symmetry is intimately related to intrinsic symmetry properties of MDMs in a quasi-2D ferrite disk. The hidden helical structure of MS-potential wave functions inside a ferrite disk gives evidence for a geometrical phase associated with the MS-wave dynamics [22, 23]. From a spectral analysis of MS-potential wave functions in a quasi-2D ferrite disk, it follows that due to special boundary conditions on a lateral surface of a ferrite disk, one has Berry connection, double-valued-function surface magnetic currents and fluxes of gauge electric fields. The MDM ferrite disk is characterized by eigen electric moments (anapole moments) [22, 23].

The paper is organized as follows. In Section 2 we present topological textures of MDM-vortex polaritons obtained from numerical simulation of a structure of a rectangular waveguide with an enclosed small ferrite disk. We analyze the scattering-matrix characteristics and give a detailed analysis of the fields for these MDM-vortex polaritons. Section 3 is devoted to an analytical consideration of the possible origin of MDM-vortex polaritons. We study the helicity and the orthogonality conditions of the MDMs in a ferrite disk and analyze properties of the observed split-state resonances. The paper is concluded by a summary in Section 4.

## 2. Distinct topological textures of MDM-vortex polaritons

As a simple model, one can consider the fields associated with MDMs in a quasi-2D ferrite disk, as the structures originated from rotating magnetic dipole and rotating electric quadrupole. Due to such field structures one can observe the power flow vortices inside a ferrite disk and in a near-field vacuum region [17 – 19]. Distinct topological textures of MDM-vortex polaritons become evident from numerical studies based on the HFSS electromagnetic simulation program (the software based on FEM method produced by ANSOFT Company). In a numerical analysis in the present paper, we use the same disk parameters as in Refs. [17 – 19]: the yttrium iron garnet (YIG) disk has a diameter of $D = 3$ mm and the disk thickness is $t = 0.05$ mm; the disk is normally magnetized by a bias magnetic field $H_0 = 4900$ Oe; the saturation magnetization of a ferrite is $4\pi M_s = 1880$ G. Similarly to Refs. [17 – 19], a ferrite disk is placed inside a $TE_{10}$-mode rectangular X-band waveguide symmetrically to its walls so that a disk axis is perpendicular to a wide wall of a waveguide. The waveguide walls are made of a perfect electric conductor (PEC). For better understanding the field structures we use a ferrite disk with a very small linewidth of $\Delta H = 0.1$ Oe. Fig. 1 shows the module and phase frequency characteristics of the reflection (the $S_{11}$ scattering-matrix parameter) coefficient, whereas Fig. 2 shows the module and phase frequency characteristics of the transmission (the $S_{21}$ scattering-matrix parameter) coefficient. The resonance modes are designated in succession by numbers $n$ = 1, 2, 3… An insertion in Fig. 1 (a) shows geometry of a structure: a ferrite disk enclosed in a rectangular waveguide.

The field structures of MDM oscillations are strongly different from the field structures of egen modes of an empty rectangular waveguide [17 – 19]. MDM-vortex polaritons appear as a result of interaction of MDM oscillations with propagating electromagnetic waves. In the represented characteristics one can clearly see that, starting from the second mode, the coupled states of electromagnetic fields with MDM vortices are split-resonance states. In Fig. 1, these split resonances are denoted by single and double primes. The split resonances are characterized by two coalescent behaviors, namely: strong transmission and strong reflection of electromagnetic waves in a waveguide. In a case of the observed strong transmission (resonances denoted by a single prime), microwave excitation energy is transformed into MDM energy and re-emitted in the forward direction, whereas in a case of strong reflection (resonances denoted by double primes), microwave excitation energy is transformed into MDM energy and re-emitted in the backward direction. As the most pronounced illustration of the MDM-vortex-polariton characteristics, we focus our study on the second-mode ($n = 2$) coalescent resonances designated as $2'$- and $2''$- resonances. The $2'$- resonance is the low-frequency resonance, while $2''$- resonance is the high-frequency resonance. Figs. 3 and 4 show the power flow density distributions in a near-field vacuum region (a vacuum plane 75 mkm above a ferrite disk) for the $2'$- and $2''$ - resonance, respectively. One can see that for the $2'$- resonance there are two power flow vortices of the near fields (above and below a ferrite disk) with opposite topological charges (positive for counterclockwise vortex and negative for clockwise vortex). Because of such a topological structure near a ferrite disk, power flow in a waveguide effectively bends around a ferrite disk resulting in a negligibly small reflected wave. Evidently, at the $2'$-resonance frequency one has electromagnetic field transparency and cloaking for a ferrite particle. Contrarily to the above behavior, at the $2''$- resonance frequency there is strong reflection of electromagnetic waves in a waveguide. The power flow distribution above and below a ferrite disk is characterized by a single-vortex behavior with strong localization of an electromagnetic field. Such a resonance behavior (the $2''$- resonance) is known from our previous studies in Refs. [17 – 19]. Figs. 5 and 6 show the electric field distributions at two time phases ($\omega t = 0°$ and $\omega t = 90°$) in a vacuum region (75 mkm above a ferrite disk) for the $2'$- and $2''$ - resonance, respectively. Evidently, there is a rotational degree of freedom for the electric-field vectors resulting in a precession behavior of the electric field in vacuum.

For understanding properties of the MDM-vortex polaritons, we should correlate the field structures in the near-field vacuum region and inside a ferrite disk. The power flow density inside a ferrite disk for the $2'$- and $2''$- resonances are shown in Figs. 7 (a) and 7 (b), respectively. Figs. 8 and 9 show the electric field distributions at two time phases ($\omega t = 0°$ and $\omega t = 90°$) inside a ferrite disk for the $2'$- and $2''$- resonances, respectively. One can see that, in spite of some small differences in pictures of the fields and power flows, the shown distributions inside a ferrite disk for the split-state $2'$- and $2''$- resonances are almost the same. There are the pictures typical for the 2$^{nd}$ MDM [18, 19]. At the same time, the near fields in vacuum are quite different for the $2'$- and $2''$- resonances (see Figs. 3 – 6).

A distinctive feature of the electric field structures, both inside and outside a ferrite disk, is the presence of local circular polarization (photon spin) of electromagnetic waves together with a cyclic evaluation of the electric field around a disk axis. An explicit illustration of such evolutions of an electric field inside a disk is given in Fig. 10 (a) in an assumption that the rotating field vector has a constant amplitude. It is evident that when (for a given radius and a certain time phase $\omega t$) an azimuth angle $\theta$ varies from 0 to $2\pi$, the electric-field vector accomplishes the $2\pi$ geometric-phase rotation. This is the nonintegrable phase factor arising from a circular closed-path parallel transport of a system (an electric field vector). We have wave plates continuously rotating locally and rotating along the power flow circulating around a disk axis. A disk axis can be considered as a line defect corresponding to "adding" or "subtracting" an angle around a line. Such a line defect, in which rotational symmetry is violated, is an example of disclination. For better understanding the fact of $2\pi$ geometric-phase rotation of the electric-field vector, in Figs. 10 (b) and 10 (c) we show, respectively, evolutions of the radial and azimuthal parts of polarization (for a given radius and a certain time phase $\omega t$). One can conclude that microwave fields of the MDM-vortex polaritons are characterized by spin and orbital angular momentums. The spin and orbital angular momentums, both oriented normally to a disk plane, are in a proper direction for the interaction. Our analysis indicates that the propagation of the waveguide-mode field is influenced by the "spin-orbit" interaction in a ferrite particle. Such a "spin-orbit" interaction plays the role of the vector potential for the waveguide-mode field. The waveguide mode travels through the substance of whirling power flow and is deflected by the MDM-vortex vector potential. The waveguide field experiences the MDM power-flow vortex in the same way as a charged-particles wave experiences a vector potential

Aharonov-Bohm effect. This is similar to the optical Aharonov-Bohm effect when light travels through the whirling liquid [24].

It is necessary to note, however, that there are some doubts as to whether spin and orbital angular momentums are, in general, separately physical observable. Maxwell's equations in vacuum are not invariant under spin and orbital angular momentums, because of the transversality condition on the electromagnetic fields. They are invariant under the total (spin plus orbital) angular momentum operator. As a consequence, no photon state exists with definite values of spin and orbital angular momentum. It is relevant here to make some comparison of our results with properties of spin and orbital angular momentums of the fields in optical helical beams and optical near-field structures. In optics, it was shown that besides the angular momentum related to photon spin, light beams in free space may also carry orbital angular momentum. Such beams are able to exert torques on matter [25]. For particles trapped on the beam axis, both spin and orbital angular momenta are transferred with the same efficiency so that the applied torque is proportional to the total angular momentum [26]. It is pointed out in literature that the situation with understanding physics of spin and orbital angular momentums of light becomes more complicated in the optical near-field regime, where the optical fields under the influence of the material environment exhibit quite different nature from those in free space. Reflecting the nature of coupled modes of optical fields and material excitations, the physical quantities associated with optical near-field interactions have distinctive properties in comparison with optical radiation in the far field [27]. These effects in optics can be useful for understanding physics of the MDM-vortex polaritons in microwaves.

For a more detailed analysis of the field polarization in MDM-vortex polaritons, we insert a piece of a metal wire inside a waveguide vacuum region. A metal rod is made of a perfect electric conductor and has very small sizes compared to free-space electromagnetic wavelength: the diameter of 200 mkm and the length of 1 mm. When such a small rod is placed rather far from a ferrite disk and oriented along an electric field of an empty rectangular waveguide, the field structure of an entire waveguide is not noticeably disturbed. At the same time, due to such a small rod one can extract the "fine structure" of the fields at MDM resonances. Fig. 11 and 12 show the electric fields on a small PEC rod for the $2'$- and $2''$- resonances, respectively. The rod position in a waveguide is shown in an insert in Fig. 11. The rod is placed along a disk axis. A gap between the lower end of the rod and the disk plane is 300 mkm. From Fig. 11, it is evident that for the $2'$- resonance (when one has electromagnetic field transparency and cloaking) there is a trivial picture of the electric field induced on a small electric dipole inside a waveguide. At the same time, from Fig. 12 one sees that in a case of the $2''$- resonance (when there is strong reflection of electromagnetic waves in a waveguide), a PEC rod behaves as a small line defect on which rotational symmetry is violated. The observed evolution of the radial part of polarization gives evidence for presence of a geometrical phase in the vacuum-region field of the MDM-vortex polariton.

In general, for the fields of MDM-vortex polaritons one has an intergo-differential problem. Because of symmetry breakings, a MDM ferrite disk, being a very small particle compared to free-space electromagnetic wavelength, is a singular point for electromagnetic fields in a waveguide. A topological character of such a singularity can be especially well illustrated by a structure of a magnetic field on a waveguide metallic wall for resonances characterizing by strong reflection and localization of electromagnetic fields in a waveguide. In the spectral characteristics shown in Fig. 1, such resonances are designated by numbers 1, 2'', 3'', 4'',… For better representation, we will consider a resonance with the most pronounced field topology – the 1-resonance. Fig. 13 shows a magnetic field on an upper wide wall of a waveguide for the 1-resonance at different time phases. One has to correlate this picture with a magnetic field nearly a ferrite disk. Fig. 14 shows a magnetic field in a vacuum region (75 mkm above a ferrite disk) for the 1- resonance at different time phases. On the waveguide metal wall, magnetic field is pure planar (2D). From Fig. 13, one clearly sees time-dependent motion of a rotating planar magnetic field is characterized by surface topological magnetic charges (STMCs). The STMCs are points of divergence and convergence of a 2D magnetic field (or a surface magnetic flux density $\vec{B}_S$) on a waveguide wall. As it is evident from Fig. 13, one has non-zero outward (inward) flows of a vector field $\vec{B}_S$ through a closed line $C$ nearly surrounding points of divergence or convergence: $\oint_C \vec{B}_S \cdot \vec{n}_S \, dC \neq 0$. Here $\vec{n}_S$ is a normal vector to contour $C$, lying on a surface of a metal wall. At the same time, it is clear, however, that $\vec{\nabla}_S \cdot \vec{B}_S = 0$, since there is zero magnetic field at points of divergence or convergence. Topological singularities on the metal waveguide wall show unusual properties. One can see that for the region bounded by the circle $C$, no planar variant of the divergence theorem takes place.

## 3. Theoretical insight into the origin of the MDM-vortex-polariton structures

A non-integrable electromagnetic problem of a ferrite disk in a rectangular waveguide, following from closed-loop nonreciprocal phase behavior on a lateral surface of a ferrite disk, can be solved numerically based on the HFSS program. From the above numerical analysis we concluded that in a thin ferrite disk, microwave fields of MDM-vortex polaritons exhibit properties which could be characterized as originated from spin and orbital angular momentums. It can be supposed that in spite of the fact that the spin and orbital angular momentums of the microwave fields are not separately observable, the shown split-resonance states of MDM-vortex polaritons are due to "spin-orbit" interactions. In general, numerical studies do not give us ability for necessary understanding physics of the MDM-vortex polaritons. At the same time, recently developed [22, 23, 28, 29] analytical approach for MDM resonances, based on a formulation of a spectral problem for quantum-like scalar wave function – the MS-potential wave function $\psi$ – may clarify physical properties of MDM-vortex polaritons. The analytical description of MDM oscillations in a quasi-2D ferrite particle rest on two cornerstones: (i) all precessing electrons in a magnetically ordered ferrite sample are described by a unique macroscopic (MS-potential) wave function $\psi$ and (ii) the phase of this wave function is well-defined over the whole ferrite-disk system, i.e. MDMs are quantum-like macroscopic states maintaining the global phase coherence. As it was shown in Refs. [17 – 19], the spectra obtained based on the $\psi$-function analysis are in a good correspondence with the numerical HFSS spectra.

The pictures of rotating (precessing) electric fields, shown in a previous section of the paper, give evidence for the left-right asymmetry of electromagnetic fields. The observed near-field photon helicity should be intimately related to hidden helical properties of MDMs. While creation of a full-wave electromagnetic field analysis of helicity in MDM-vortex polaritons entails great difficulties (because of nonintegrability, i.e. path-dependence of the problem), analytical solutions of the $\psi$–function spectral problem can explain hidden helical properties of MDM resonances. Because of nonreciprocal phase behavior on a lateral surface of a ferrite disk, the phase variations for resonant $\psi$ functions are both in azimuth, $\theta$, and axial, $z$, directions. This shows that proper spectral problem solutions for MDMs should be obtained in a helical coordinate system [22, 30]. The helices are topologically nontrivial structures and the phase relationships for waves propagating in such structures could be very special. Unlike the Cartesian or cylindrical coordinate systems, the helical coordinate system is not orthogonal and separate the right-handed and left-handed solutions are admitted. Since the helical coordinates are nonorthogonal and curvilinear, different types of helical coordinate systems can be suggested. In an analysis of the MS-wave propagation in a helical coordinate system in Refs. [22, 30], the Waldron's coordinate system was used [31].

In the Waldron helical system $(r,\phi,\zeta)$, the Walker equation for MS-potential wave function [11] has a form [22, 30]:

$$\frac{\partial^2 \psi}{\partial r^2} + \frac{1}{r}\frac{\partial \psi}{\partial r} + \frac{1}{r^2}\frac{\partial^2 \psi}{\partial \phi^2} + \left(\frac{1}{\mu} + \tan^2 \alpha_0\right)\frac{\partial^2 \psi}{\partial \zeta^2} - 2\frac{1}{r}\left(\tan \alpha_0\right)^{(R,L)} \frac{\partial^2 \psi}{\partial \phi\, \partial \zeta} = 0, \qquad (1)$$

where superscripts $R$ and $L$ mean, respectively, right-handed and left-handed helical coordinate systems, and $\mu$ is a diagonal component of the permeability tensor. For pitch $p$, the pitch angles are defined from the relations:

$$\left(\tan \alpha_0\right)^{(R)} \equiv \tan \alpha_0 \equiv \bar{p}/r \quad \text{and} \quad \left(\tan \alpha_0\right)^{(L)} = -\tan \alpha_0 = -\bar{p}/r, \qquad (2)$$

where $\bar{p} = p/2\pi$. The quantities $\tan \alpha_0$ and $\bar{p}$ are assumed to be positive. For a given direction of a bias magnetic field, there are four helical modes. Inside a ferrite disk of radius $\Re$ ($r \leq \Re$) the solutions are in the form:

$$\begin{aligned}
\psi^{(1)} &= a_1 J_{(w-\bar{p}\beta)}\left[(-\mu)^{1/2} \beta\, r\right] e^{-iw\phi} e^{-i\beta\zeta}, \\
\psi^{(2)} &= a_2 J_{(w-\bar{p}\beta)}\left[(-\mu)^{1/2} \beta\, r\right] e^{+iw\phi} e^{-i\beta\zeta}, \\
\psi^{(3)} &= a_3 J_{(w-\bar{p}\beta)}\left[(-\mu)^{1/2} \beta\, r\right] e^{+iw\phi} e^{+i\beta\zeta}, \\
\psi^{(4)} &= a_4 J_{(w-\bar{p}\beta)}\left[(-\mu)^{1/2} \beta\, r\right] e^{-iw\phi} e^{+i\beta\zeta}.
\end{aligned} \qquad (3)$$

For an outside region ($r \geq \Re$) one has:

$$\begin{aligned}
\psi^{(1)} &= b_1 K_{(w-\bar{p}\beta)}(\beta\, r)\, e^{-iw\phi} e^{-i\beta\zeta}, \\
\psi^{(2)} &= b_2 K_{(w-\bar{p}\beta)}(\beta\, r)\, e^{+iw\phi} e^{-i\beta\zeta}, \\
\psi^{(3)} &= b_3 K_{(w-\bar{p}\beta)}(\beta\, r)\, e^{+iw\phi} e^{+i\beta\zeta}, \\
\psi^{(4)} &= b_4 K_{(w-\bar{p}\beta)}(\beta\, r)\, e^{-iw\phi} e^{+i\beta\zeta}.
\end{aligned} \qquad (4)$$

Here $J$ and $K$ are Bessel functions of real and imaginary arguments, respectively. Coefficients $a_{1,2,3,4}$ and $b_{1,2,3,4}$ are amplitude coefficients. It was shown [22, 30] that for a given direction of a bias magnetic field (oriented along a disk axis) there exist two types of double-helix resonances in a quasi-2D ferrite disk. One resonance state is specified by the $\psi^{(1)} \leftrightarrow \psi^{(4)}$ phase correlation, when a closed-loop phase way is due to equalities for the wave numbers: $w^{(1)} = w^{(4)}$ and $\beta^{(1)} = \beta^{(4)}$. Another resonance state is specified by the $\psi^{(2)} \leftrightarrow \psi^{(3)}$ phase correlation with a closed-loop phase way due to equalities for the wave numbers: $w^{(2)} = w^{(3)}$ and $\beta^{(2)} = \beta^{(3)}$. The resonance $\psi^{(1)} \leftrightarrow \psi^{(4)}$, being characterized by the right-hand rotation (with respect to a bias magnetic field directed along an axis of a disk, the $z$-axis) of a composition of helices, is conventionally called as the (+) resonance. The resonance $\psi^{(2)} \leftrightarrow \psi^{(3)}$, with the left-hand rotation of a helix composition, is conventionally called as the (−) resonance. In a case of the (+) double-helix resonance, the azimuth phase over-running of the MS-potential wave functions is in a correspondence with the right-hand resonance rotation of magnetization in a ferrite magnetized along z-axis by a DC magnetic field [11].

The helical-mode resonances of lossless magneto-dipole oscillations in a ferrite disk are not characterized by the orthogonality relations. It can be shown that for two helices giving a double-helix resonance in a ferrite disk there are deferent power flow densities [30]. Moreover, for such modes there are no properties of parity ($\mathcal{P}$) and time-reversal ($\mathcal{T}$) invariance – the $\mathcal{PT}$-invariance. Solutions in Eqs. (3) and (4), being multiplied by a time factor $e^{i\omega t}$, describe propagating helical waves. Inversion of a direction of a bias magnetic field gives inversion of time and so inversion of a sign of the off-diagonal component of the permeability tensor [11]. From an analysis in Ref. [22], one can see that for the contrarily directed bias magnetic field, the (+) double-helix resonance appears due to the $\psi^{(2)} \leftrightarrow \psi^{(3)}$ phase correlated helices, while the (−) double-helix resonance is due to the $\psi^{(1)} \leftrightarrow \psi^{(4)}$ interference. Such time inversion, however, cannot be accompanied by the space reflection with respect to a disk plane. Because of lack of reflection symmetry for helical modes, there is no mutual reflection for helical modes $\psi^{(1)}$ and $\psi^{(3)}$, as well as no mutual reflection for helical modes $\psi^{(4)}$ and $\psi^{(2)}$. The $\mathcal{PT}$-symmetry breaking does not guarantee real-eigenvalue spectra, but, in a case of a lossless structure, can give spectra with pairs of complex-conjugate eigenvalues. It was shown, however, that by virtue of quasi-two dimensionality of the problem, one can reduce solutions from helical to cylindrical coordinates with proper separation of variables [22]. This gives integrable solutions for MDMs in a cylindrical coordinate system. As we discuss below, such solutions can be considered as $\mathcal{PT}$-invariant. The $\mathcal{PT}$-invariance properties of MDMs in a quasi-2D ferrite disk play an essential role in physics of the observed topologically distinctive states.

For the (+) double-helix resonance, one can introduce the notion of an effective membrane function $\tilde{\varphi}$ and describe the spectral problem by a differential-matrix equation [19, 22, 23, 28, 29]

$$\left( \hat{L}_\perp - i\beta \, \hat{R} \right) \tilde{V} = 0 , \tag{5}$$

where $\tilde{V} \equiv \begin{pmatrix} \tilde{\vec{B}} \\ \tilde{\varphi} \end{pmatrix}$, $\tilde{\varphi}$ is a dimensionless membrane MS-potential wave function and $\tilde{\vec{B}}$ is a dimensionless membrane function of a magnetic flux density. In Eq. (5), $\hat{L}_\perp$ is a differential-matrix operator:

$$\hat{L}_\perp \equiv \begin{pmatrix} (\tilde{\mu}_\perp)^{-1} & \nabla_\perp \\ -\nabla_\perp \cdot & 0 \end{pmatrix}, \tag{6}$$

where subscript $\perp$ means correspondence with the in-plane, $r, \theta$, coordinates, $\beta$ is the MS-wave propagation constant along z axis ($\psi = \tilde{\varphi} \, e^{-i\beta z}$, $\vec{B} = \tilde{\vec{B}} \, e^{-i\beta z}$), $\tilde{\mu}$ is the permeability tensor, $\hat{R}$ is a matrix: $\hat{R} \equiv \begin{pmatrix} 0 & \vec{e}_z \\ -\vec{e}_z & 0 \end{pmatrix}$, and $\vec{e}_z$ is a unit vector along z axis. The boundary condition of continuity of a radial component of the magnetic flux density on a lateral surface of a ferrite disk of radius $\Re$ is expressed as [22, 23, 28]:

$$\mu\left(\frac{\partial \tilde{\varphi}}{\partial r}\right)_{r=\Re^-} - \left(\frac{\partial \tilde{\varphi}}{\partial r}\right)_{r=\Re^+} = -i\frac{\mu_a}{\Re}\left(\frac{\partial \tilde{\varphi}}{\partial \theta}\right)_{r=\Re^-}, \qquad (7)$$

where $\mu$ and $\mu_a$ are, respectively, diagonal and off-diagonal components of the permeability tensor $\ddot{\mu}$. The modes described by a differential-matrix equation (5), are conventionally called as *L*-modes. With use of separation of variables and boundary conditions of continuity of a MS-potential wave function and a magnetic flux density on disk surfaces, one obtains solutions for *L*-modes. For a ferrite disk of radius $\Re$ and thickness *d*, the solutions are [18, 22]:

$$\psi(r,\theta,z,t) = C\xi(z)J_\nu\left(\frac{\beta r}{\sqrt{-\mu}}\right)e^{-i\nu\theta}e^{i\omega t} \qquad (8)$$

inside a ferrite disk ($r \leq \Re$, $-d/2 \leq z \leq d/2$) and

$$\psi(r,\theta,z,t) = C\xi(z)K_\nu(\beta r)\,e^{-i\nu\theta}e^{i\omega t} \qquad (9)$$

outside a ferrite disk (for $r \geq \Re$, $-d/2 \leq z \leq d/2$). In these equations, $\nu$ is an azimuth number, $J_\nu$ and $K_\nu$ are the Bessel functions of order $\nu$ for real and imaginary arguments, *C* is a dimensional coefficient, and $\xi(z)$ is an amplitude factor. For the solutions represented by Eqs. (8), (9), the characteristic equation (7) has a form:

$$(-\mu)^{1/2}\left(\frac{J'_\nu}{J_\nu}\right)_{r=\Re} + \left(\frac{K'_\nu}{K_\nu}\right)_{r=\Re} - \frac{\nu \mu_a}{\beta \Re} = 0, \qquad (10)$$

where the prime denotes differentiation with respect to the argument. It is necessary to note that, in accordance with the (+) double-helix-resonance conditions [22], the azimuth number $\nu$ takes only integer and positive quantities. The membrane MS-potential functions $\tilde{\varphi}$ for *L* modes do not have the standing-wave configuration in a disk plane, but are azimuthally-propagating waves. For rotationally non-symmetric waves one has the azimuth power flow density [18]:

$$\left(p_q(r,z)\right)_\theta = \frac{\tilde{\varphi}_q^*(r)}{8\pi}\omega\, C_q^2\left(\xi_q(z)\right)^2\left[-\tilde{\varphi}_q(r)\frac{\mu}{r}\nu - \mu_a\frac{\partial \tilde{\varphi}_q(r)}{\partial r}\right]. \qquad (11)$$

Here *q* is a number of radial variations (for a given azimuth number $\nu$). Since an amplitude of a MS-potential function is equal to zero at $r=0$, the power-flow-density is zero at the disk center. Eq. (11) describes the power-flow-density vortex inside a ferrite disk. For the 2$^{nd}$ MDM, the circulating power-flow density was analytically calculated based on Eq. (11) and with use of the disk parameters mentioned above for the HFSS simulation. Fig. 7 (c) shows the results of such a calculation for $\nu = +1$ and $q = 2$. One can see that this analytical representation is in a good correlation with the numerical results of the power-flow densities inside a ferrite disk for the 2′- and 2″- resonances [see Figs. 7 (a) and (b)].

The spectral-problem solutions for *L* modes give real eigenvalues of propagation constants $\beta$ and orthogonality conditions for eigenfunctions $\begin{pmatrix} \tilde{\vec{B}} \\ \tilde{\varphi} \end{pmatrix}$. For a certain mode *n*, the norm is defined as [18, 19, 23, 28]

$$N_n = \int_S \left( \tilde{\varphi}_n \tilde{\vec{B}}_n^* - \tilde{\varphi}_n^* \tilde{\vec{B}}_n \right) \cdot \vec{e}_z dS, \qquad (12)$$

where *S* – a square of an open MS-wave cylindrical waveguide. The norm $N_n$, being multiplied by a proper dimensional coefficient, corresponds to the power flow density of the waveguide mode *n* through a waveguide cross section. In an assumption of separation of variables in a cylindrical coordinate system, this power flow along *z*-axis should be considered independently of the azimuth power flow density defined by Eq. (11). It follows, however, that because of special symmetry properties of *L* modes (azimuthal non-symmetry of membrane functions $\tilde{\varphi}$), representation of the norm by Eq. (12) is not so definite and operator $\hat{L}_\perp$ cannot be considered as self-adjoint operator. At the same time, as we will show, operator $\hat{L}_\perp$ is $\mathcal{PT}$-invariant:

$$\hat{L}_\perp = \left( \hat{L}_\perp \right)^{\mathcal{PT}}. \qquad (13)$$

This may presume the absence of complex eigenvalues for *L* modes [32].

It is worth beginning our studies of the symmetry properties with some illustrative analysis of modes in a disk resonator. Let us consider, initially, a simple case of a non-magnetic disk resonator. For resonance processes in a non-ferrite dielectric resonator characterizing by an oscillation period *T*, time shifts $t = 0 \rightarrow t = T$ and $t = T \rightarrow t = 0$ are formally equivalent since there is no chosen direction of time for the electron motion processes inside a dielectric material. When a resonator has cylindrical geometry, a counterclockwise rotating wave (RW) acquires the phase $\Phi = 2\pi k$ ($k = 1, 2, 3, ...$) at the time shift $t = 0 \rightarrow t = T$ and at the time shift $t = T \rightarrow t = 0$, a clockwise RW acquires the same phase, $\Phi = 2\pi k$. Since in a cylindrically symmetric non-ferrite resonator, dynamical behaviors are not distinguished by time inversion, a counterclockwise RW cannot be excited separately from a clockwise RW. Let us consider now a ferrite-disk resonator. Suppose that for a given direction of a normal bias magnetic field $H_0$ there is a counterclockwise rotating wave (RW) in a ferrite disk and this wave acquires phase $\Phi_1$ at the time shift $t = 0 \rightarrow t = T$, where *T* is an oscillation period. Making time inversion (inversion of a direction of a bias magnetic field $H_0$), we obtain a clockwise RW. For this wave, we then consider the time shift $t = -T \rightarrow t = 0$. We suppose that in this case a clockwise RW acquires phase $\Phi_2$. A system comes back to its initial state when both partial rotating processes (counterclockwise and clockwise RWs), with phases $\Phi_1$ and $\Phi_2$, are involved. Since, geometrically, a system is azimuthally symmetric, it is evident that $|\Phi_1| = |\Phi_2| \equiv \Phi$. A total minimal phase due to two RW processes should be equal to $2\pi$. Generally, one has $\Phi_1 + \Phi_2 = 2\Phi = 2\pi k$ or

$$\Phi = k\pi, \qquad (14)$$

where *k* = 1, 2, 3, … The phases for RWs in a MS-mode cylindrical resonator are shown in Fig. 15. To bring a system to its initial state one should involve the time reversal operations. When only one direction of a normal bias magnetic field is given and quantities *k* are odd integers, the MS wave rotating in a certain azimuth direction (either counterclockwise or clockwise) should make two rotations around a disk axis to come back to its initial state. It means that for a given direction of a bias magnetic field and for odd integer *k*, a membrane function $\tilde{\varphi}$ behaves as a double-valued function. It is worth noting that, in general, the phase of the final state differs from that of the initial state by

$$\Phi = \Phi_d + \Phi_g, \qquad (15)$$

where $\Phi_d$ and $\Phi_g$ are the dynamical and geometric phases, respectively. If only the topology of the path is altered, then only $\Phi_g$ varies [33]. In our case, this fact is illustrated very clearly by Figs. 8 and 9. Let us compare the positions of electric field vectors in Figs. 8 and 9 for a certain dynamical phase $\omega t = 0°$, for example. One can see that these vectors are shifted in space at angle of 90°. At the same time, since the frequency shift between the 2′- and 2″- resonances is negligibly small ($\Delta f/f \approx 13/8645 = 0.0015$), the dynamical phase ($\omega t = 0°$) is the same. The observed strong variation of a geometric phase against a background on a non-varying dynamical phase is a good confirmation of a topological character of the split-resonance states.

Let us analyze now properties of operator $\hat{L}_\perp$. Following a standard way of solving boundary problems in mathematical physics [34, 35], one can consider two joint boundary problems: the main boundary problem and the conjugate boundary problem. Both problems are described by differential equations which are similar to Eq. (5). The main boundary problem is expressed by a differential equation $\left(\hat{L}_\perp - i\beta \hat{R}\right) \tilde{V} = 0$ and the conjugate boundary problem is expressed by an equation: $\left(\hat{L}_\perp^\circ - i\beta^\circ \hat{R}\right) \tilde{V}^\circ = 0$. From a formal point of view, it is supposed initially that these are different equations: there are different differential operators, different eigenfunctions and different eigenvalues. A form of differential operator $\hat{L}_\perp^\circ$ one gets from integration by parts:

$$\int_S (\hat{L}_\perp \tilde{V})(\tilde{V}^\circ)^* dS = \int_S \tilde{V}(\hat{L}_\perp^\circ \tilde{V}^\circ)^* dS + \oint_{\mathcal{L}} P(\tilde{V}, \tilde{V}^\circ) \, d\ell, \qquad (16)$$

where $\mathcal{L} = 2\pi\Re$ is a contour surrounding a cylindrical ferrite core and $P(\tilde{V}, \tilde{V}^\circ)$ is a bilinear form. Operator $\hat{L}_\perp$ can be considered as a self-adjoint (Hermitian) operator when permeability tensor $\ddot{\mu}$ is a Hermitian tensor, functions $\tilde{V}$ and $\tilde{V}^\circ$ are two mutually complex conjugate functions, and a contour integral in the right-hand side of Eq. (16) is equal to zero [34, 35]. The last condition means that for an open ferrite structure [a core ferrite region (*F*) is surrounded by a dielectric region (*D*)] there are homogeneous boundary conditions for functions $\tilde{V}$ and $\tilde{V}^\circ$:

$$\oint_{\mathcal{L}} P(\tilde{V}, \tilde{V}^\circ) \, d\ell \equiv \oint_{\mathcal{L}} [P^{(F)}(\tilde{V}, \tilde{V}^\circ) + P^{(D)}(\tilde{V}, \tilde{V}^\circ)] \, d\ell = 0. \qquad (17)$$

As we will show, in a general case, $\tilde{V}$ and $(\tilde{V}^\circ)^*$ are not two mutually complex conjugate functions and so we do not have self-adjointness of operator $\hat{L}_\perp$. At the same time, there exist necessary conditions for the $\mathcal{PT}$-invariant homogeneous boundary conditions (17) resulting in the $\mathcal{PT}$-invariance of operator $\hat{L}_\perp$.

For the contour integral in the right-hand side of Eq. (16), we have [23]

$$\oint_{\mathcal{L}} P(\tilde{V},\tilde{V}^\circ)\, d\ell = -\oint_{\mathcal{L}} M(\tilde{V},\tilde{V}^\circ)\, d\ell - \oint_{\mathcal{L}} N(\tilde{V},\tilde{V}^\circ)\, d\ell, \tag{18}$$

where

$$\oint_{\mathcal{L}} M(\tilde{V},\tilde{V}^\circ)\, d\ell \equiv \oint_{\mathcal{L}} \left\{ \left[ \mu \left( \frac{\partial \tilde{\varphi}}{\partial r} \right)_{r=\mathfrak{R}^-} - \left( \frac{\partial \tilde{\varphi}}{\partial r} \right)_{r=\mathfrak{R}^+} \right] (\tilde{\varphi}^\circ)^*_{r=\mathfrak{R}} \right. \\ \left. - (\tilde{\varphi})_{r=\mathfrak{R}} \left[ \mu \left( \frac{\partial \tilde{\varphi}^\circ}{\partial r} \right)_{r=\mathfrak{R}^-} - \left( \frac{\partial \tilde{\varphi}^\circ}{\partial r} \right)_{r=\mathfrak{R}^+} \right]^* \right\} d\ell \tag{19}$$

and

$$\oint_{\mathcal{L}} N(\tilde{V},\tilde{V}^\circ)\, d\ell \equiv \oint_{\mathcal{L}} \left[ \left( i\mu_a \frac{\partial \tilde{\varphi}}{\partial \theta} \right)(\tilde{\varphi}^\circ)^* - (\tilde{\varphi})\left( \left( i\mu_a \frac{\partial \tilde{\varphi}^\circ}{\partial \theta} \right)^\circ \right)^* \right]_{r=\mathfrak{R}} d\ell. \tag{20}$$

In these equations we used expressions for radial components of the magnetic flux density: (a) for the main boundary problem, $\tilde{B}_r = -\left( \mu \frac{\partial \tilde{\varphi}}{\partial r} + i\mu_a \frac{\partial \tilde{\varphi}}{\partial \theta} \right)$ in a ferrite region and $\tilde{B}_r = -\frac{\partial \tilde{\varphi}}{\partial r}$ in a dielectric, and (b) for the conjugate boundary problem, $\tilde{B}_r^\circ = -\left[ \mu \frac{\partial \tilde{\varphi}^\circ}{\partial r} + \left( i\mu_a \frac{\partial \tilde{\varphi}}{\partial \theta} \right)^\circ \right]$ in a ferrite region and $\tilde{B}_r^\circ = -\frac{\partial \tilde{\varphi}^\circ}{\partial r}$ in a dielectric.

Following Eq. (15), we represent $\tilde{\varphi}$ and $\tilde{\varphi}^\circ$ as

$$\tilde{\varphi} \equiv \delta \tilde{\eta}, \tag{21}$$

$$\tilde{\varphi}^\circ \equiv \delta^\circ \tilde{\eta}^\circ. \tag{22}$$

Functions $\tilde{\eta} \sim e^{-i\vartheta_d}$ and $\tilde{\eta}^\circ \sim e^{-i\vartheta_d^\circ}$ are characterized by dynamical phases $\vartheta_d$ and $\vartheta_d^\circ$, respectively, while functions $\tilde{\delta} \sim e^{-i\vartheta_g}$ and $\tilde{\delta}^\circ \sim e^{-i\vartheta_g^\circ}$ are characterized by geometrical phases $\vartheta_g$ and $\vartheta_g^\circ$, respectively. Evidently, double-valuedness of functions $\tilde{\varphi}$ and $\tilde{\varphi}^\circ$ is due to the presence of geometrical phases. It means that functions $\tilde{\eta}$ and $\tilde{\eta}^\circ$ are single-valued functions while functions $\tilde{\delta}$ and $\tilde{\delta}^\circ$ are double-valued functions. With such a representation, we can say also that functions $\tilde{\eta}$ and $\tilde{\eta}^\circ$ are described by "orbital coordinates", whereas functions $\tilde{\delta}$ and $\tilde{\delta}^\circ$ are described by "spinning coordinates". Since $\tilde{\eta}$ and $\tilde{\eta}^\circ$ are space-reversally invariant, functions $\tilde{\eta}$ and $\left(\tilde{\eta}^\circ\right)^*$ can be considered just as complex conjugate functions. At the same time, $\tilde{\delta}$ and $\left(\tilde{\delta}^\circ\right)^*$ are not complex conjugate functions.

To satisfy homogeneous boundary relation (17), we consider conditions when both $\oint_\mathcal{L} M(\tilde{V},\tilde{V}^\circ)\,d\ell$ and $\oint_\mathcal{L} N(\tilde{V},\tilde{V}^\circ)\,d\ell$ are equal to zero. For integral $\oint_\mathcal{L} N(\tilde{V},\tilde{V}^\circ)\,d\ell$, expressed by Eq. (20), there is possibility to analyze the orthogonality conditions separately for the "orbital coordinates" and "spinning coordinates" [23, 36]. The "orbital coordinate" integral has a form:

$$\oint_\mathcal{L}\left[\left(i\mu_a \frac{\partial \tilde{\eta}}{\partial \theta}\right)(\tilde{\eta}^\circ)^* - (\tilde{\eta})\left(\left(i\mu_a \frac{\partial \tilde{\eta}}{\partial \theta}\right)^\circ\right)^*\right]_{r=\Re} d\ell \qquad (23)$$

and the "spinning coordinate" integral is expressed as:

$$\oint_\mathcal{L}\left[\left(i\mu_a \frac{\partial \tilde{\delta}}{\partial \theta}\right)(\tilde{\delta}^\circ)^* - (\tilde{\delta})\left(\left(i\mu_a \frac{\partial \tilde{\delta}}{\partial \theta}\right)^\circ\right)^*\right]_{r=\Re} d\ell. \qquad (24)$$

For integral (23) one has

$$\oint_\mathcal{L}\left[\left(i\mu_a \frac{\partial \tilde{\eta}}{\partial \theta}\right)(\tilde{\eta}^\circ)^* - (\tilde{\eta})\left(\left(i\mu_a \frac{\partial \tilde{\eta}}{\partial \theta}\right)^\circ\right)^*\right]_{r=\Re} d\ell = \oint_\mathcal{L}\left[\left(i\mu_a \frac{\partial \tilde{\eta}}{\partial \theta}\right)(\tilde{\eta})^* - (\tilde{\eta})\left(i\mu_a \frac{\partial \tilde{\eta}}{\partial \theta}\right)^*\right]_{r=\Re} d\ell \equiv 0. \qquad (25)$$

At the same time, just only with time inversion one cannot have integral (24) equal to zero. Function $\tilde{\delta}^\circ$ is considered as the $\mathcal{P}$-transformed function with respect to function $\tilde{\delta}$ and to have integral (24) equal to zero, one has to consider the combined $\mathcal{PT}$ transformation. It is evident that in a quasi-2D ferrite-disk structure, geometrical-phase circular running waves $\tilde{\delta}$ will have opposite direction of rotation for $z$-axis reflection, so $\mathcal{P}$ is classified as space reflection with respect to $z$-axis. When one considers waves $\left(\tilde{\delta}^\circ\right)^*$ as the waves, which are $\mathcal{PT}$ transformed relatively to waves $\tilde{\delta}$, one has integral (24) equal to zero. In general, $\left(\tilde{\varphi}^\circ\right)^*$ is considered as $\mathcal{PT}$ transformed function relatively to $\tilde{\varphi}$. Because of $\mathcal{PT}$ invariance of function $\tilde{\varphi}$, one concludes that integral (19) is identically equal to zero. As a result, one has zero integral (18).

An introduction of membrane functions in a quasi-2D ferrite-disk structure allows reducing the problem of parity transformation to one-dimension reflection in space. Geometrical-phase circular running of membrane function $\tilde{\varphi}$ will have opposite direction of rotation for $z$-axis reflection. From the above analysis of the contour integral $\oint_\mathcal{L} P(\tilde{V},\tilde{V}^\circ)\,d\ell$ it follows that study of the $\mathcal{PT}$ invariance of operator $\hat{L}_\perp$ with eigenfunctions $\begin{pmatrix} \tilde{\vec{B}} \\ \tilde{\varphi} \end{pmatrix}$ can be reduced to an analysis of $\mathcal{PT}$ properties of membrane functions $\tilde{\varphi}$ on a lateral surface of a ferrite disk. Certainly, only the equation for boundary conditions reflects a nonreciprocal phase behavior and so path-dependence of the boundary-value problem. It is clear that simultaneous change of a sign of $\mu_a$ (the time reversal) and a sign of derivative $\left(\frac{\partial \tilde{\varphi}}{\partial \theta}\right)_{r=\Re^-}$ (the space reflection) in the right-hand side of Eq. (7) leaves this equation invariable. This is evidence for the $\mathcal{PT}$- invariance. For a value of a MS-potential wave function on a lateral surface of a ferrite disk, $\tilde{\varphi}|_{r=\Re}$, we can write

$$\mathcal{PT}\tilde{\varphi}|_{r=\Re}(z) = \tilde{\varphi}^*|_{r=\Re}(-z) = \tilde{\varphi}|_{r=\Re}(z). \tag{26}$$

There is also the possibility to introduce the orthogonality relation for two modes:

$$\left(\tilde{\varphi}_p|_{r=\Re}, \tilde{\varphi}_q|_{r=\Re}\right) = \oint_{\mathcal{L}} \left(\tilde{\varphi}_p|_{r=\Re}(z)\right)\tilde{\varphi}_q^*|_{r=\Re}(-z)d\ell = \oint_{\mathcal{L}} \left(\tilde{\varphi}_q|_{r=\Re}(z)\right)\left[\mathcal{PT}\tilde{\varphi}_p|_{r=\Re}(z)\right]d\ell. \tag{27}$$

Here we assume that the spectrum under consideration is real and contour $\mathcal{L}$ is a real line. This orthogonality relation has different meanings for even and odd quantities $k$ in Eq. (14). For even quantities $k$ in Eq. (14), the edge waves show reciprocal phase behavior for propagation in both azimuthal directions. Contrarily, for odd quantities $k$ in Eq. (14), the edge waves propagate only in one direction of the azimuth coordinate. In the case of even $k$, the orthogonality relation (27) can be written as $\left(\tilde{\varphi}_p|_{r=\Re}, \tilde{\varphi}_q|_{r=\Re}\right) = \delta_{pq}$, where $\delta_{mn}$ is Kronecker delta. With respect to this relation, for odd $k$ one has $\left(\tilde{\varphi}_p|_{r=\Re}, \tilde{\varphi}_q|_{r=\Re}\right) = -\delta_{pq}$. In a general form, the inner product (27) can be written as

$$\left(\tilde{\varphi}_p|_{r=\Re}, \tilde{\varphi}_q|_{r=\Re}\right) = (-1)^k \delta_{pq}. \tag{28}$$

We are faced with a fact that in the bound states for functions $\tilde{\varphi}|_{r=\Re}$ there are equal numbers of positive-norm and negative-norm states. To some extent, our results resemble the results of the $\mathcal{PT}$-symmetry studies in quantum mechanics structures with a complex Hamiltonian [32, 37, 38]. Similarly to paper [32], we can introduce a certain operator $\hat{\mathcal{C}}$, which is the observable that represents the measurement of the signature of the $\mathcal{PT}$ norm of a state. While, in the problem under consideration, one has quasi-orthogonality of $L$ modes and pseudo-hermicity of operator $\hat{L}_\perp$, there should exist a certain operator $\hat{\mathcal{C}}$ that action of $\hat{\mathcal{C}}$ together with the $\mathcal{PT}$ transformation will give the hermicity condition and real-quantity energy eigenstates. A form of operator $\hat{\mathcal{C}}$ is found from an assumption that the operator $\hat{\mathcal{C}}$ acts only on the boundary conditions of the $L$-mode spectral problem. Such a technique was used in Refs. [23, 36].

The operator $\hat{\mathcal{C}}$ is a special differential operator in a form of $i\frac{\mu_a}{\Re}\left(\vec{\nabla}_\theta\right)_{r=\Re}$. Here $\left(\vec{\nabla}_\theta\right)_{r=\Re}$ is the spinning-coordinate gradient. It means that for a given direction of a bias field, operator $\hat{\mathcal{C}}$ acts only for one-direction azimuth variation. The eigenfunctions of operator $\hat{\mathcal{C}}$ are double-valued border functions [23, 36]. This operator allows making transformation from essential boundary conditions of *L* modes, expressed by Eqs. (7) and (10), to natural boundary conditions of so-called *G* modes, which have forms, respectively [19, 23, 28]:

$$\mu\left(\frac{\partial \tilde{\varphi}}{\partial r}\right)_{r=\Re^-} - \left(\frac{\partial \tilde{\varphi}}{\partial r}\right)_{r=\Re^+} = 0 \qquad (29)$$

and

$$(-\mu)^{1/2}\left(\frac{J'_\nu}{J_\nu}\right)_{r=\Re} + \left(\frac{K'_\nu}{K_\nu}\right)_{r=\Re} = 0 . \qquad (30)$$

The membrane functions of *G* modes are related to the orbital-coordinate system. It is evident that the quantum-like *G*-mode spectra cannot be shown by the HFSS numerical simulation. Using operator $\hat{\mathcal{C}}$ we construct the new inner product structure for boundary functions:

$$(\tilde{\varphi}_n|_{r=\Re}, \tilde{\varphi}_m|_{r=\Re}) = \oint_{\mathcal{L}}\left[\hat{\mathcal{C}}\mathcal{PT}\tilde{\varphi}_m|_{r=\Re}(z)\right](\tilde{\varphi}_n|_{r=\Re}(z))d\ell . \qquad (31)$$

As a result, one has the energy eigenstate spectrum of MS-mode oscillations with topological phases accumulated by the double-valued border functions [23]. The topological effects become apparent through the integral fluxes of the pseudo-electric fields. There are positive and negative fluxes corresponding to the counterclockwise and clockwise edge-function chiral rotations. For an observer in a laboratory frame, we have two oppositely directed anapole moments $\vec{a}^e$. This anapole moment is determined by a term $i\frac{\mu_a}{\Re}\left(\frac{\partial \tilde{\varphi}}{\partial \theta}\right)_{r=\Re^-}$ in the right-hand side of Eq. (7). For a given direction of a bias magnetic field, we have two cases $\vec{a}^e \cdot \vec{H}_0 > 0$ and $\vec{a}^e \cdot \vec{H}_0 < 0$. As it was supposed [23], there should be observed the magnetoelectric energy splitting which is, in fact, the splitting due to spin-orbit interaction.

The numerically observed topologically distinctive split-resonance states of the MDM-vortex-polariton structures are due to the $\mathcal{PT}$-invariance properties of operator $\hat{L}_\perp$. Such properties are evident, in particular, from strong variation of a geometric phase against a background on a non-varying dynamical phase. Contrarily to quasi-orthogonality of *L* modes, for *G* modes one has the hermicity condition and real-quantity energy eigenstates. Based on the above analysis, one can conclude that frequency differences for peak positions of the analytically derived *L* and *G* modes should be the same order of magnitudes as frequency differences for split-resonance states observed in the HFSS spectrum. Fig. 16, showing spectrum peak positions for the HFSS simulation and analytical *L*- and *G*-modes, gives evidence for this statement.

## 4. Conclusion

In recent years, we are witnesses of a resurgence of interest in spin wave excitations motivated by their possible use as information carriers (see Ref. [39] and references therein). Novel technological opportunities lend new momentum to the study of fundamental properties of spin-wave oscillations and interaction of these oscillations with electromagnetic fields. Among different types of microwave magnetic materials, the yttrium iron garnet (YIG) is considered as one of the most attractive material due its uniquely low magnetic damping. This ferrimagnet has the narrowest known line of ferromagnetic resonance (FMR), which results in a magnon lifetime of a few hundred nanoseconds.

Small ferrite-disk particles with MS oscillations are characterized by topologically distinctive long-living resonances with symmetry breakings. Scattering of electromagnetic fields from such small particles gives the topological states splitting. While for incident electromagnetic wave there is no difference between the left and right, in the fields scattered by a MDM ferrite particle one should distinguish left from right. It was shown that due to MDM vortices in small thin-film ferrite disks there is strong magnon-photon coupling. The coupled states – MDM-vortex polaritons – are characterized by helicity behaviors. For topologically distinctive structures of MDM-vortex polaritons one has or localization, or cloaking of electromagnetic fields. The observed strong variation of a geometric phase against a background on a non-varying dynamical phase is a good confirmation of a topological character of the split-resonance states.

Interaction of microwave fields with MDM vortices open a perspective for creating novel electromagnetic structures with unique symmetry properties. The shown properties of MDMs in a quasi-2D ferrite disk offer a particularly fertile ground in which $\mathcal{PT}$-related concepts can be realized and investigated experimentally. The important reasons for this: (a) the formal equivalence between the quantum mechanical Schrödinger equation and the *G*-mode MS wave equation [23]; (b) the possibility to manipulate the ferrite-disk geometrical and material parameters and the bias magnetic field. One of the examples of novel MDM-polariton structures could be $\mathcal{PT}$ metamaterials. There is also another interesting aspect. Since MDM vortices are topologically stable objects, they can be used as long-living microwave memory elements.

**Figure captions**

Fig. 1. Frequency characteristics of a module (a) and a phase (b) of the reflection coefficient for a rectangular waveguide with an enclosed thin-film ferrite disk. The resonance modes are designated in succession by numbers $n$ = 1, 2, 3… The coalescent resonances are denoted by single and double primes. An insertion in (a) shows geometry of a structure.

Fig. 2. Frequency characteristics of a module (a) and a phase (b) of the transmission coefficient for a rectangular waveguide with an enclosed thin-film ferrite disk. The resonance modes are designated in succession by numbers $n$ = 1, 2, 3…

Fig. 3. The power flow density distributions in a near-field vacuum region (75 mkm above a ferrite disk) for the $2'$- resonance ($f$ =8.641 GHz). (a) A general picture in a waveguide; (b) a detailed picture near a region of a ferrite disk. There is evidence for electromagnetic field transparency and cloaking.

Fig. 4. The power flow density distributions in a near-field vacuum region (75 mkm above a ferrite disk) for the $2''$ - resonance ($f$ =8.652 GHz). (a) A general picture in a waveguide; (b) a detailed picture near a region of a ferrite disk. One has strong localization and strong reflection of electromagnetic fields in a waveguide.

Fig. 5. The electric field distributions in a vacuum region (75 mkm above a ferrite disk) for the $2'$- resonance at different time phases. (a) and (b) top views; (c) and (d) side views.

Fig. 6. The electric field distributions in a vacuum region (75 mkm above a ferrite disk) for the $2''$ - resonance at different time phases. (a) and (b) top views; (c) and (d) side views.

Fig. 7. The power flow density inside a ferrite disk. Numerically obtained vortices for the $2'$ - resonance (a) and the $2''$ - resonance (b); (c) gives an analytical result for the 2nd mode obtained from Eq. (11) for $\nu = 1$; blue arrows make more precise directions of power flows.

Fig. 8. The electric field distributions at different time phases inside a ferrite disk for the $2'$ - resonance.

Fig. 9. The electric field distributions at different time phases inside a ferrite disk for the $2''$ - resonance.

Fig. 10. (a) Explicit illustration of cyclic evolutions of an electric field inside a disk in an assumption that the rotating field vector has constant amplitude. When (for a given radius and a certain time phase $\omega t$) an azimuth angle $\theta$ varies from 0 to $2\pi$, the electric-field vector accomplishes the $2\pi$ geometric-phase rotation. (b) and (c) show, respectively, evolutions of the radial and azimuthal parts of polarization (for a given radius and a certain time phase $\omega t$). One can conclude that microwave fields of the MDM-vortex polaritons are characterized by spin and orbital angular momentums. In the pictures, $\mathfrak{R}$ is a radius of a ferrite disk.

Fig. 11. Electric field on a small PEC rod for the $2'$ - resonance at different time phases. There is a trivial picture of the fields of a small electric dipole inside a waveguide. An insertion shows position of a PEC rod in a waveguide.

Fig. 12. Electric field on a small PEC rod for the $2''$ - resonance at different time phases. A PEC rod behaves as a small line defect on which rotational symmetry is violated. The observed evolution of the radial part of polarization gives evidence for presence of a geometrical phase in the vacuum-region field of the MDM-vortex polariton.

Fig. 13. Magnetic field on a wide waveguide wall for the 1- resonance at different time phases. Points A and B are, respectively, positive and negative surface topological magnetic charges.

Fig. 14. Magnetic field in a vacuum region (75 mkm above a ferrite disk) for the 1- resonance at different time phases.

Fig. 15. The phases for counterclockwise and clockwise rotating waves (RW) in a MS-mode cylindrical resonator.

Fig. 16. The spectrum peak positions for the HFSS simulation and analytical *L*- and *G*-modes. Frequency differences for peak positions of the analytically derived *L* and *G* modes are at the same order of magnitudes as frequency differences for split-resonance states observed in the HFSS spectrum.

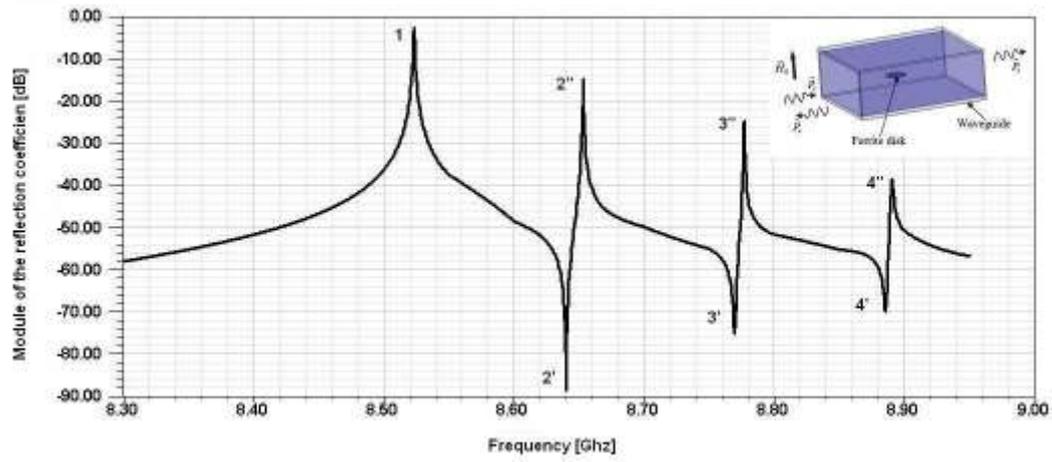

(a)

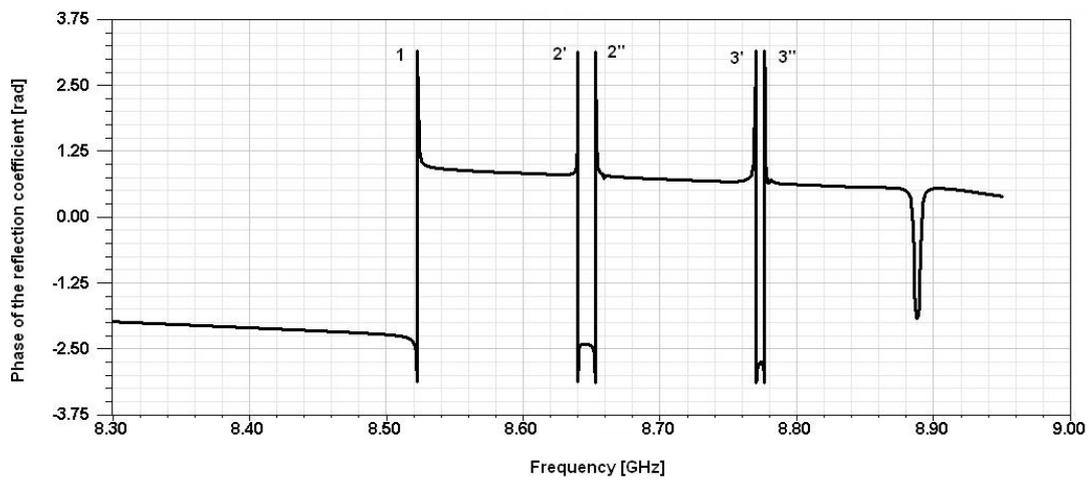

(b)

Fig. 1. Frequency characteristics of a module (a) and a phase (b) of the reflection coefficient for a rectangular waveguide with an enclosed thin-film ferrite disk. The resonance modes are designated in succession by numbers $n = 1, 2, 3…$ The coalescent resonances are denoted by single and double primes. An insertion in (a) shows geometry of a structure.

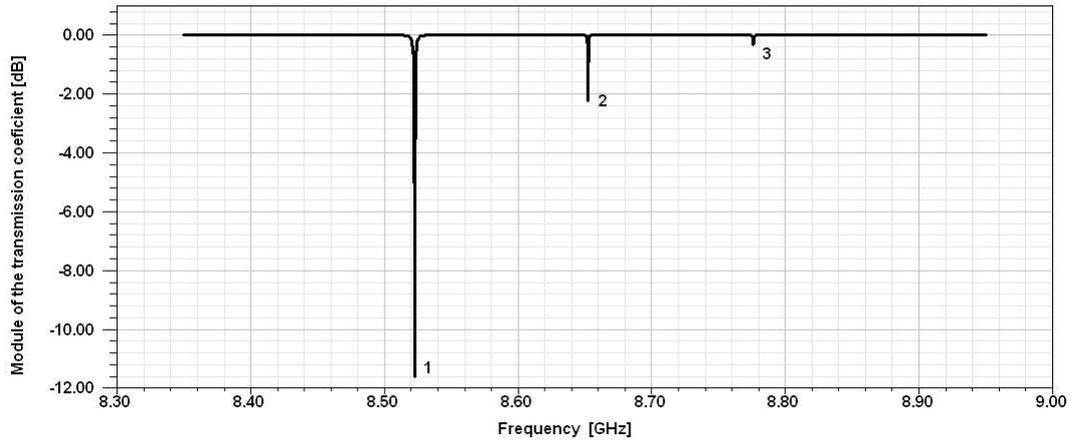

(a)

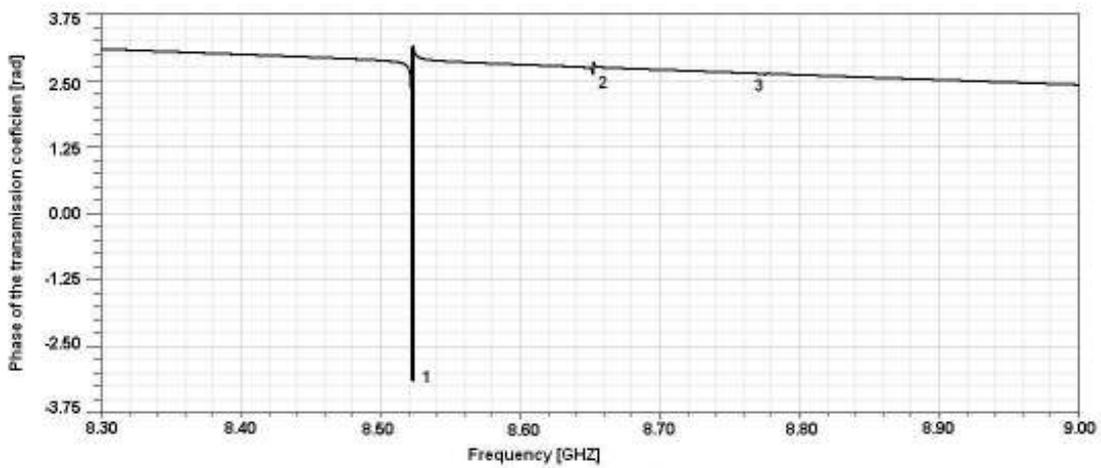

(b)

Fig. 2. Frequency characteristics of a module (a) and a phase (b) of the transmission coefficient for a rectangular waveguide with an enclosed thin-film ferrite disk. The resonance modes are designated in succession by numbers $n = 1, 2, 3…$

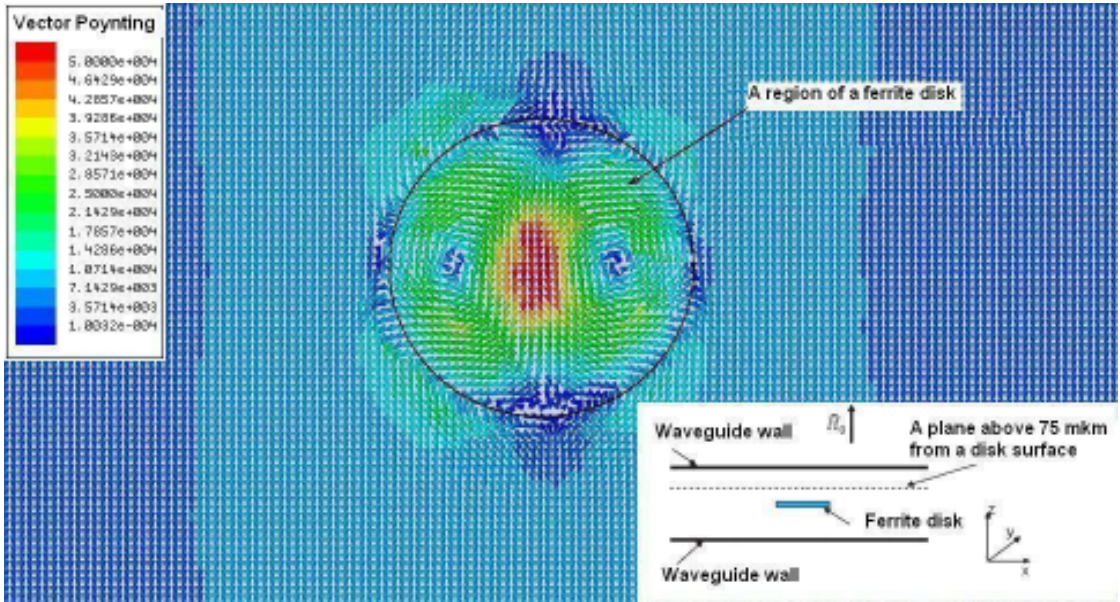

(a)

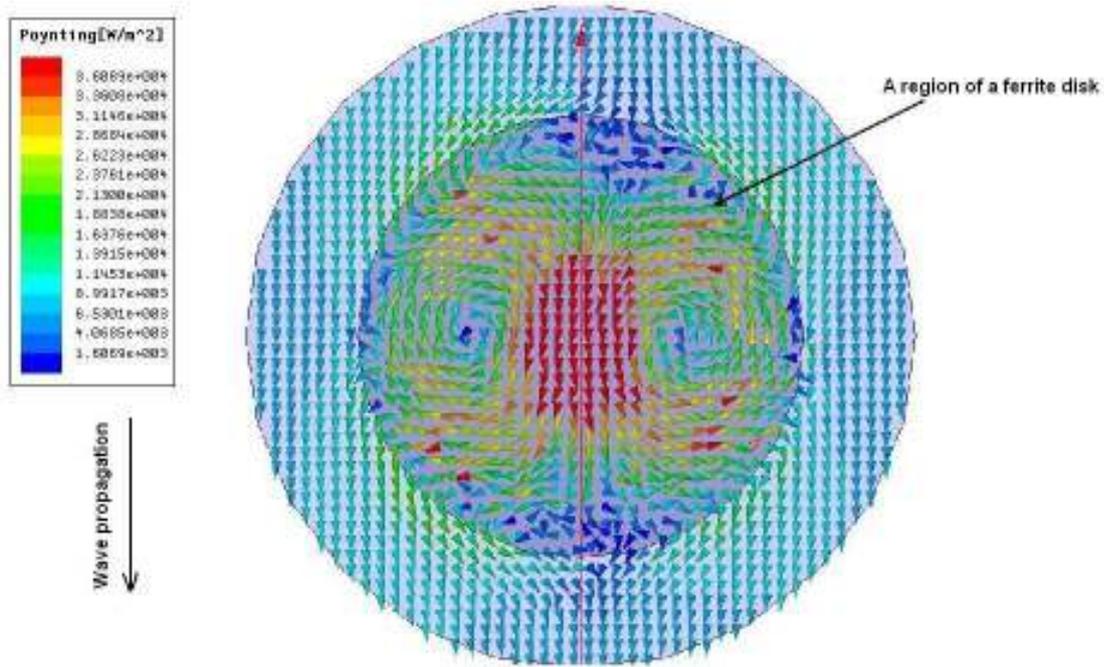

(b)

Fig. 3. The power flow density distributions in a near-field vacuum region (75 mkm above a ferrite disk) for the $2'$- resonance ($f$ =8.641 GHz). (a) A general picture in a waveguide; (b) a detailed picture near a region of a ferrite disk. There is evidence for electromagnetic field transparency and cloaking.

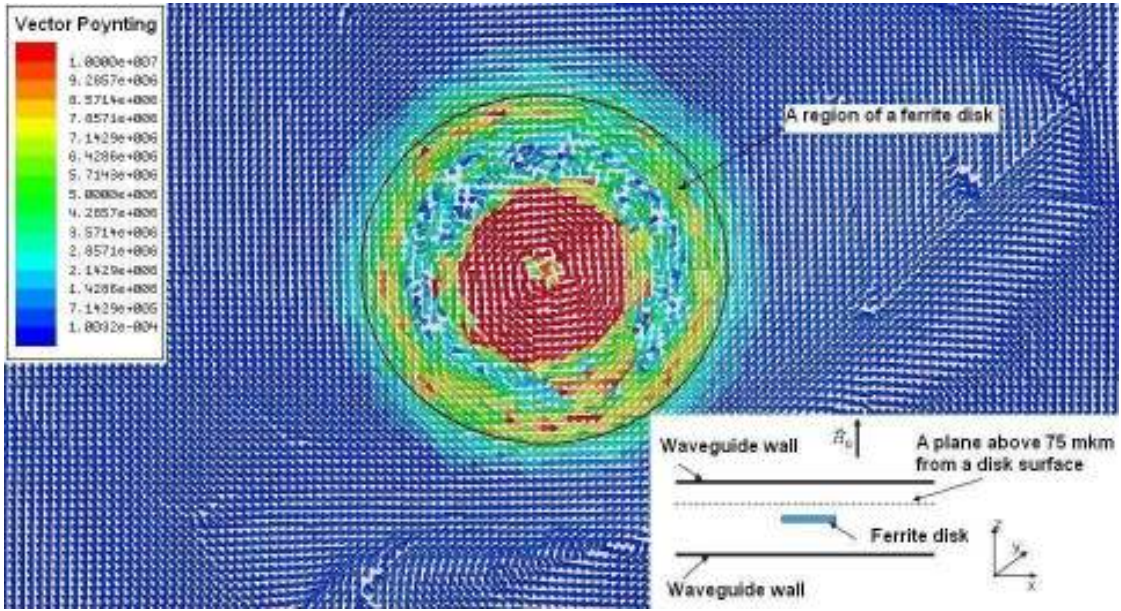

(a)

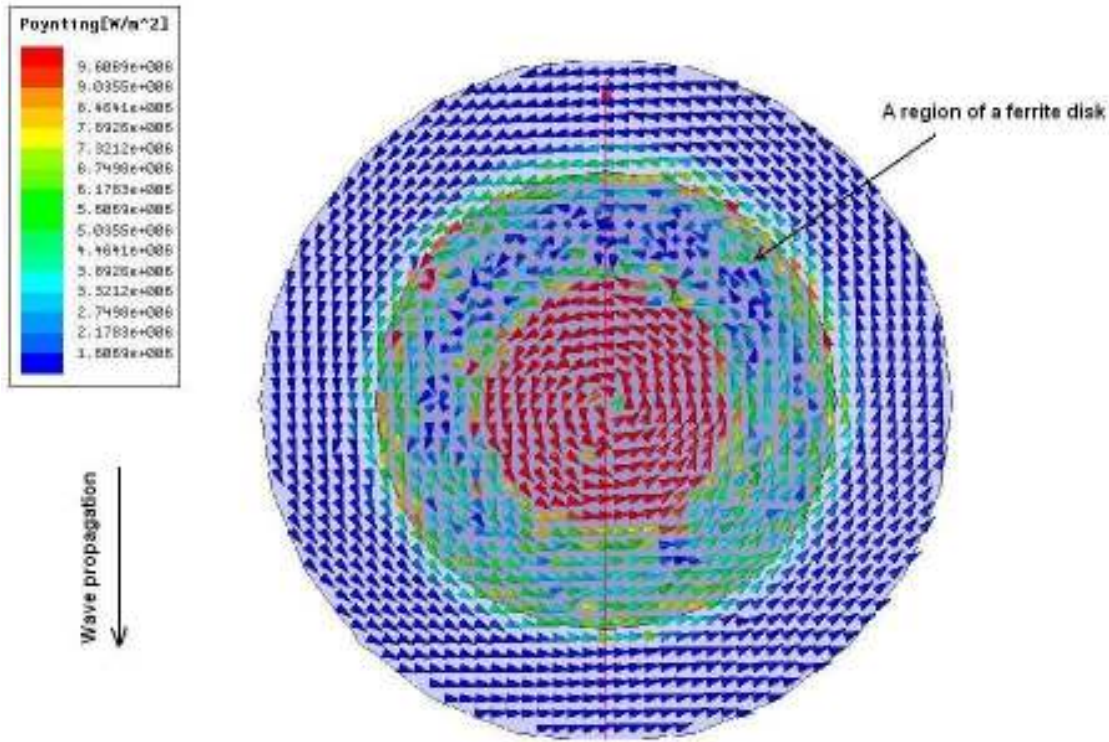

(b)

Fig. 4. The power flow density distributions in a near-field vacuum region (75 mkm above a ferrite disk) for the $2''$ - resonance ($f$ =8.652 GHz ). (a) A general picture in a waveguide; (b) a detailed picture near a region of a ferrite disk. One has strong localization and strong reflection of electromagnetic fields in a waveguide.

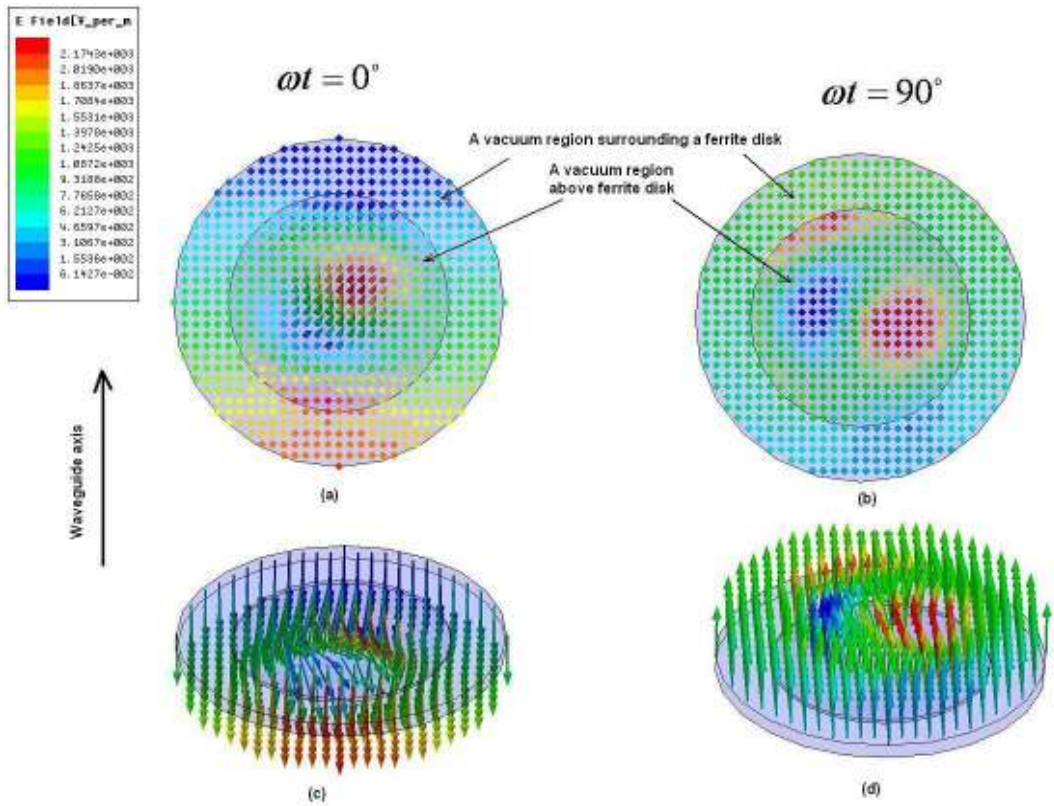

Fig. 5. The electric field distributions in a vacuum region (75 mkm above a ferrite disk) for the $2'$ - resonance at different time phases. (a) and (b) top views; (c) and (d) side views.

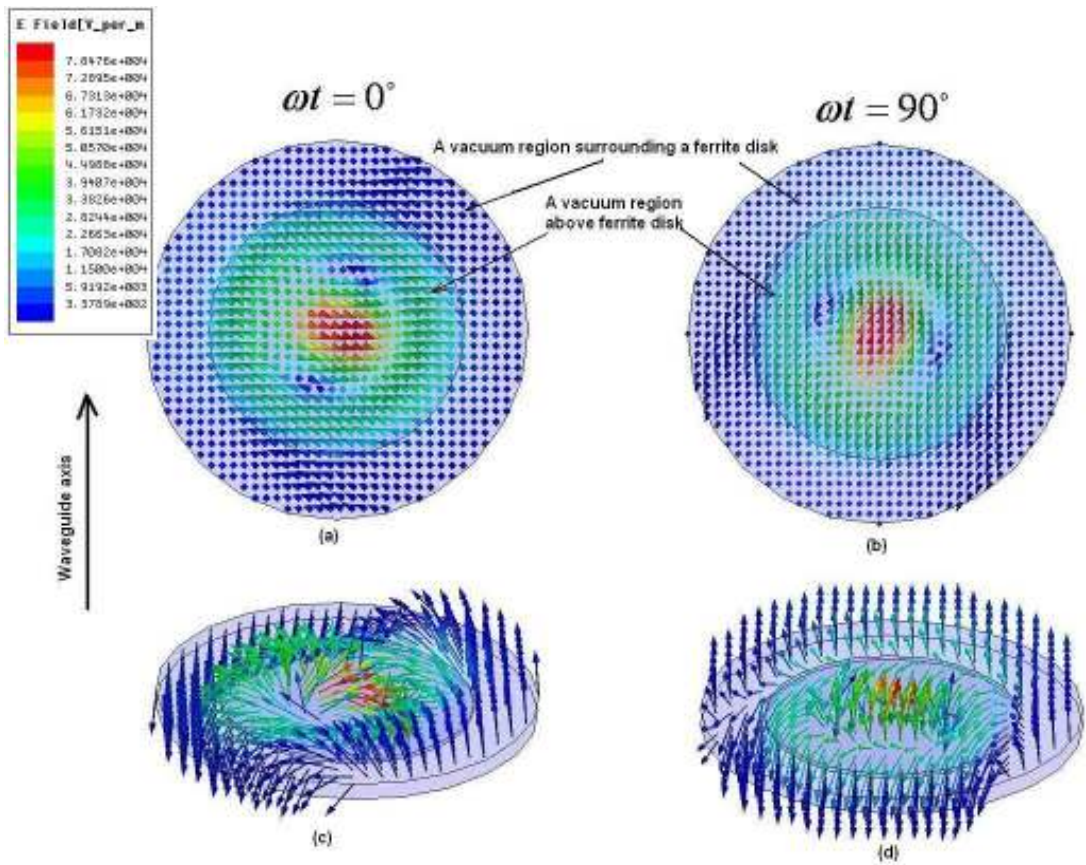

Fig. 6. The electric field distributions in a vacuum region (75 mkm above a ferrite disk) for the $2''$ - resonance at different time phases. (a) and (b) top views; (c) and (d) side views.

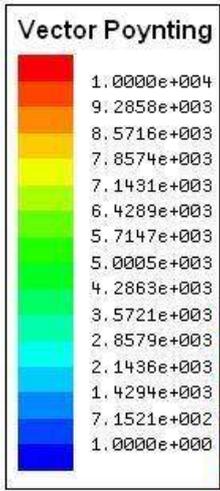
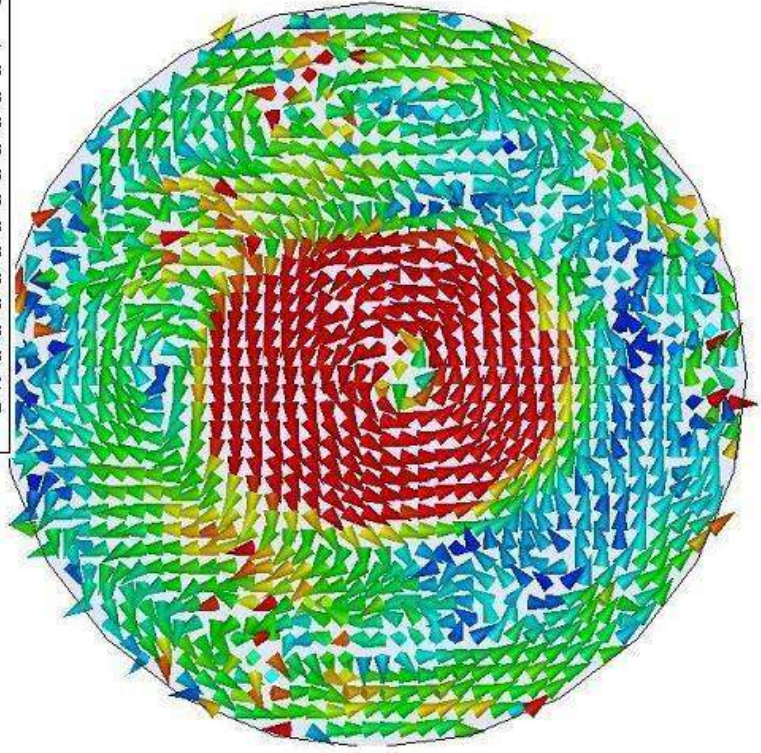

(a)

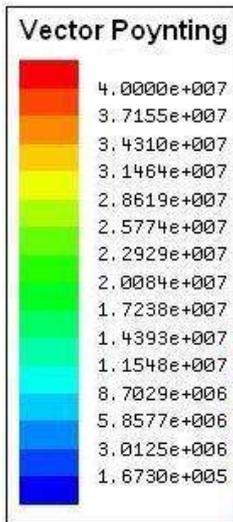
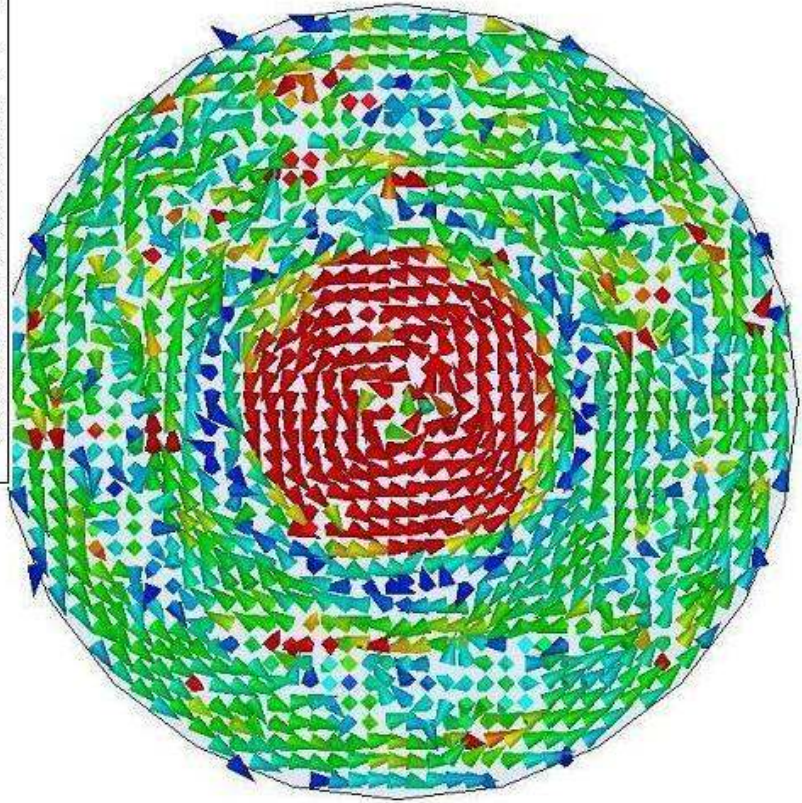

(b)

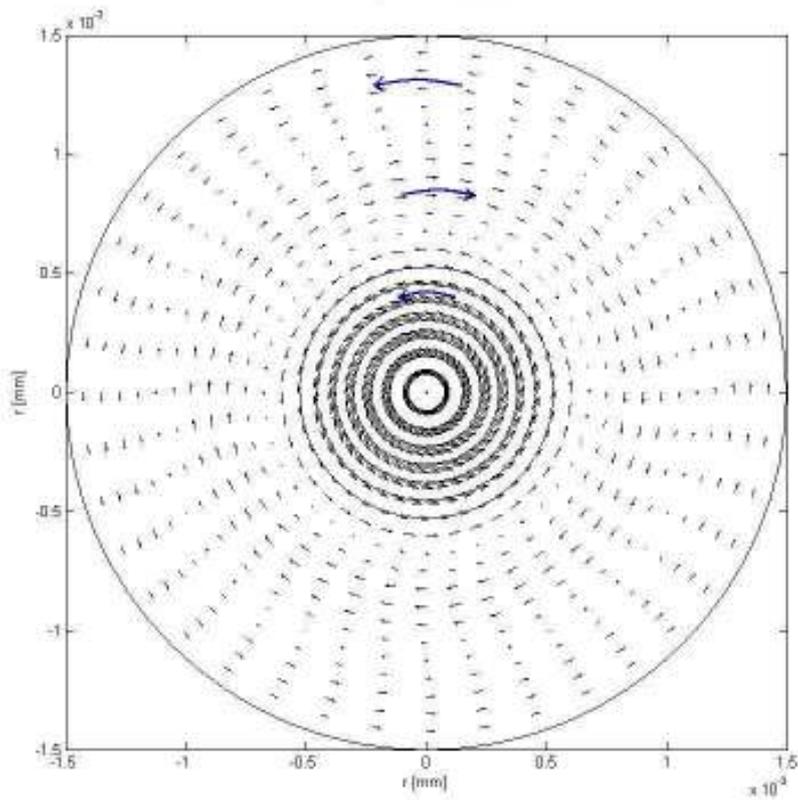

(c)

Fig. 7. The power flow density inside a ferrite disk. Numerically obtained vortices for the $2'$-resonance (a) and the $2''$- resonance (b). Fig. (c) gives an analytical result for the $2^{nd}$ mode obtained from Eq. (11) for $\nu = 1$; blue arrows make more precise directions of power flows.

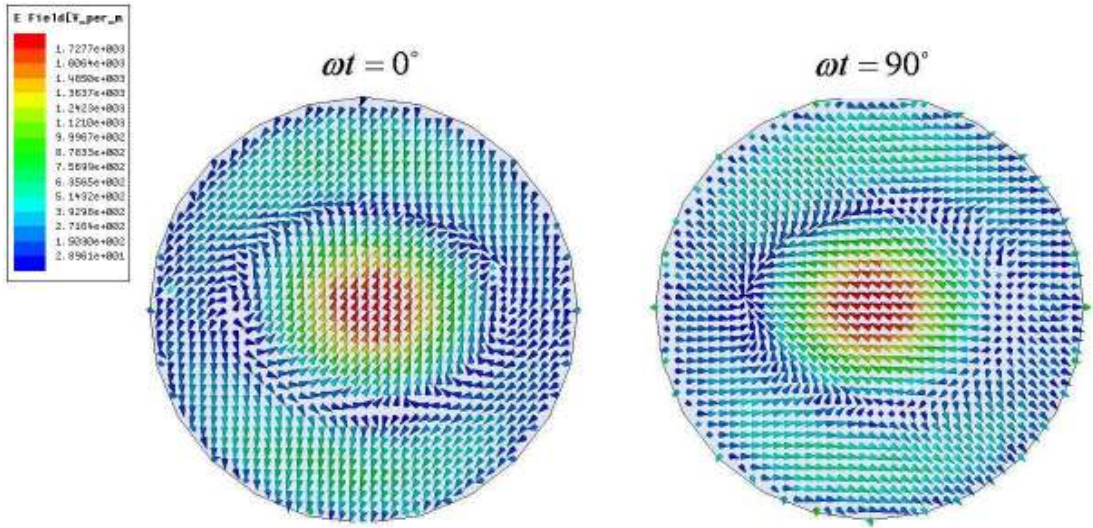

Fig. 8. The electric field distributions at different time phases inside a ferrite disk for the $2'$-resonance.

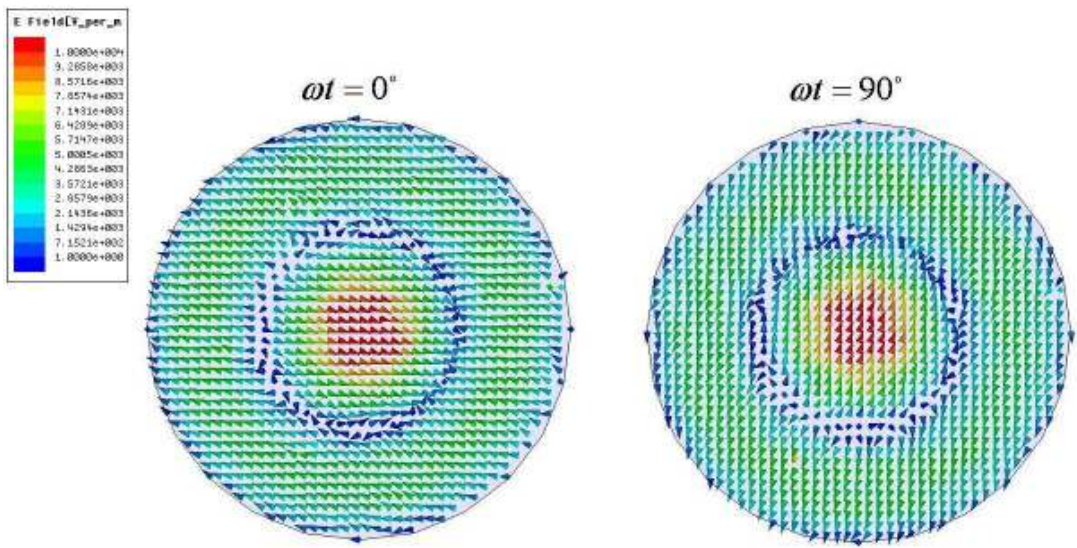

Fig. 9. The electric field distributions at different time phases inside a ferrite disk for the $2''$-resonance.

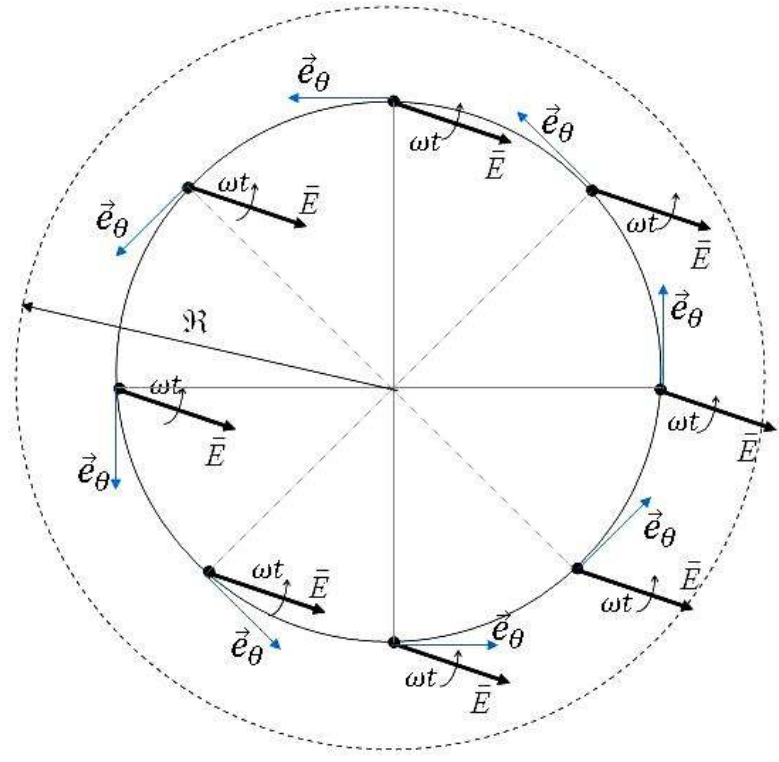

(a)

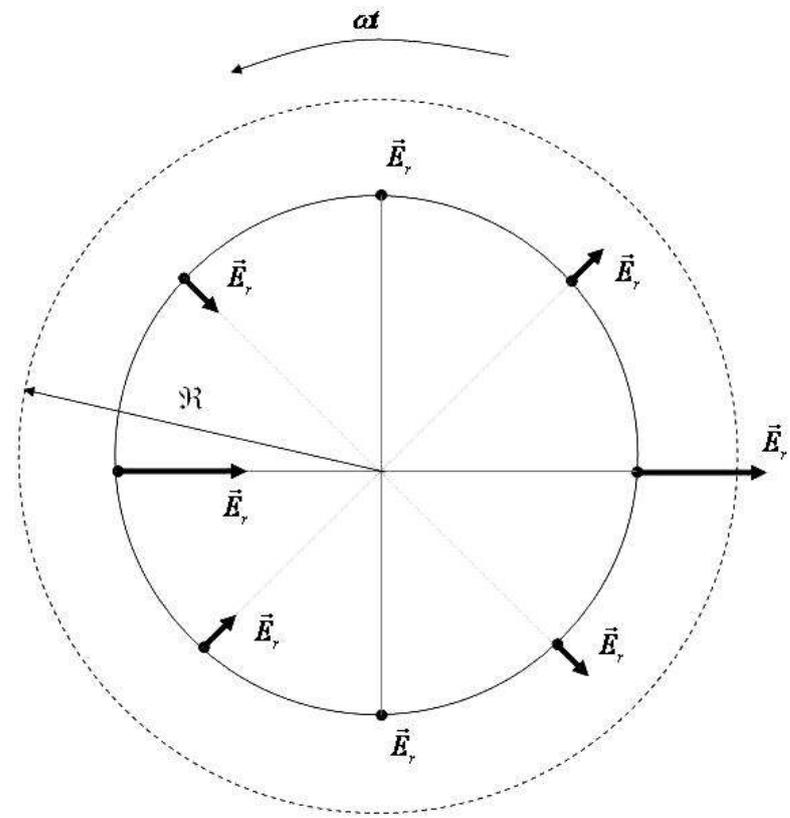

(b)

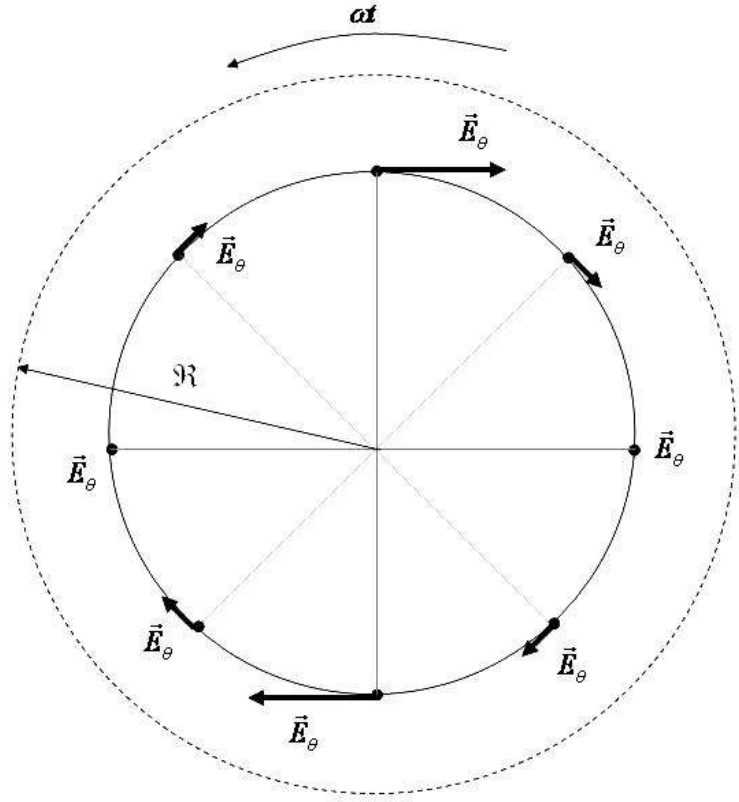

(c)

Fig. 10. (a) Explicit illustration of cyclic evolutions of an electric field inside a disk in an assumption that the rotating field vector has constant amplitude. When (for a given radius and a certain time phase $\omega t$) an azimuth angle $\theta$ varies from 0 to $2\pi$, the electric-field vector accomplishes the $2\pi$ geometric-phase rotation. (b) and (c) show, respectively, evolutions of the radial and azimuthal parts of polarization (for a given radius and a certain time phase $\omega t$). One can conclude that microwave fields of the MDM-vortex polaritons are characterized by spin and orbital angular momentums. In the pictures, $\Re$ is a radius of a ferrite disk.

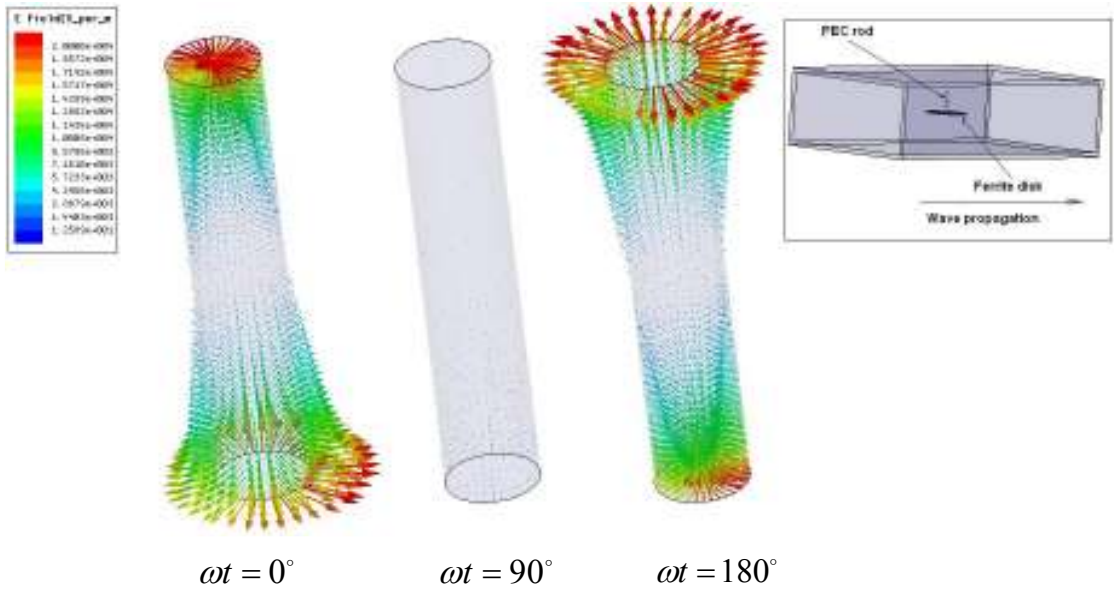

Fig. 11. Electric field on a small PEC rod for the $2'$ - resonance at different time phases. There is a trivial picture of the electric field induced on a small electric dipole inside a waveguide. An insertion shows position of a PEC rod in a waveguide.

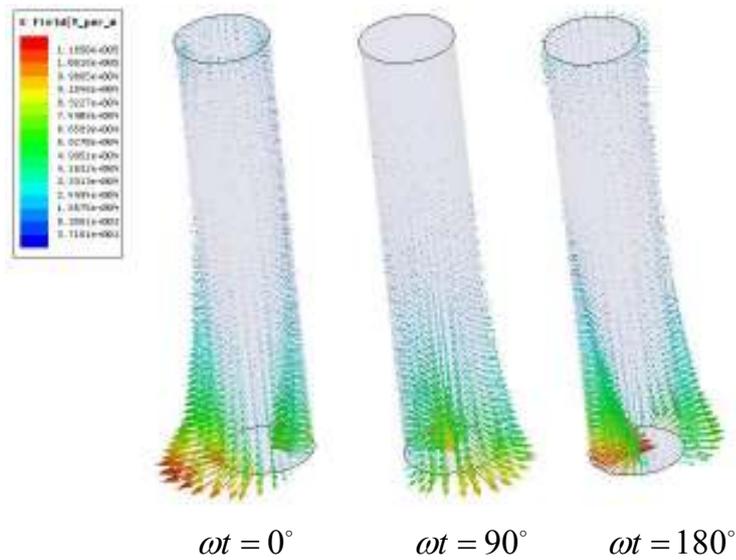

Fig. 12. Electric field on a small PEC rod for the $2''$ - resonance at different time phases. A PEC rod behaves as a small line defect on which rotational symmetry is violated. The observed evolution of the radial part of polarization gives evidence for presence of a geometrical phase in the vacuum-region field of the MDM-vortex polariton.

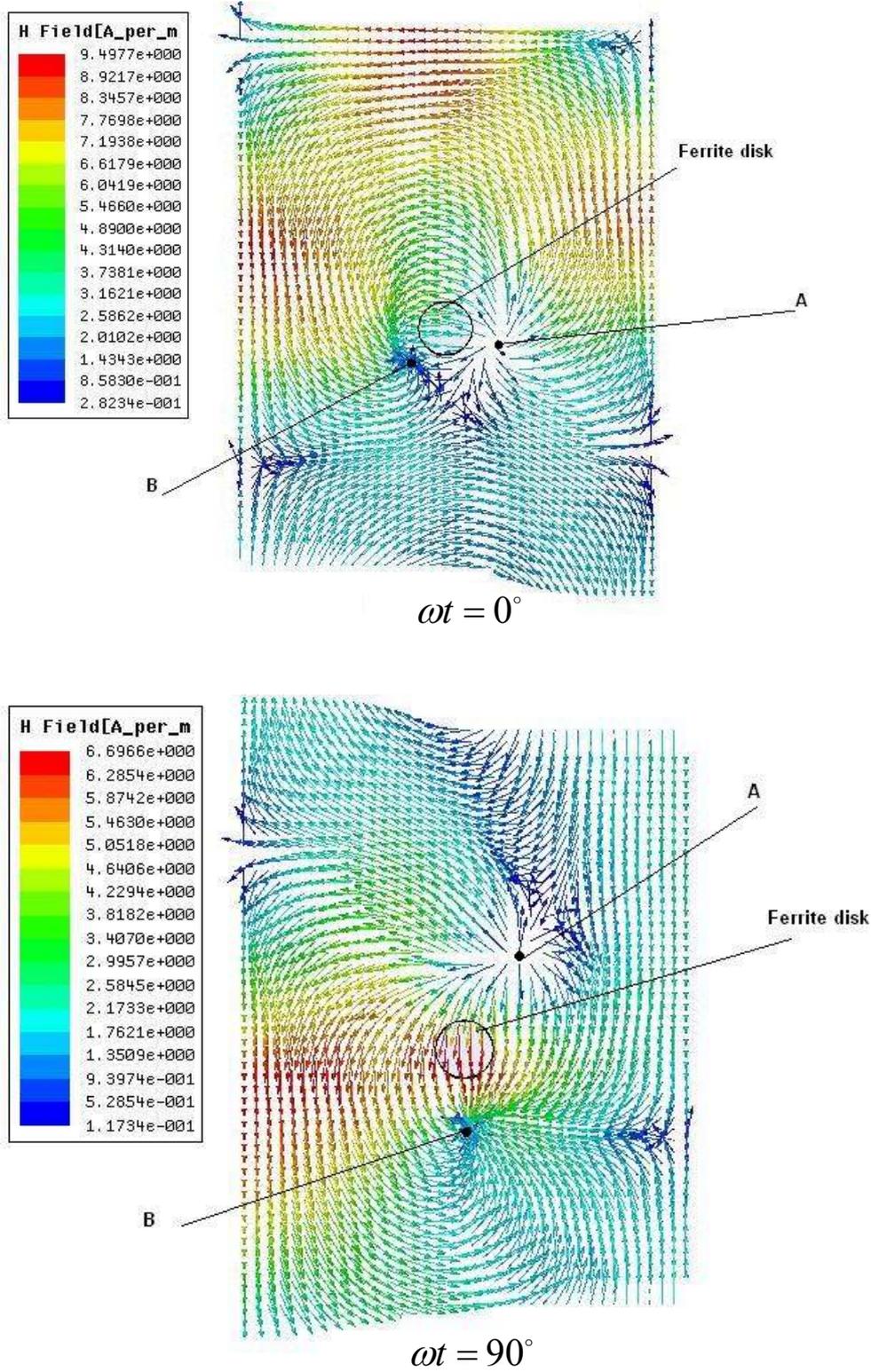

Fig. 13. Magnetic field on a wide waveguide wall for the 1- resonance at different time phases. Points A and B are, respectively, positive and negative surface topological magnetic charges.

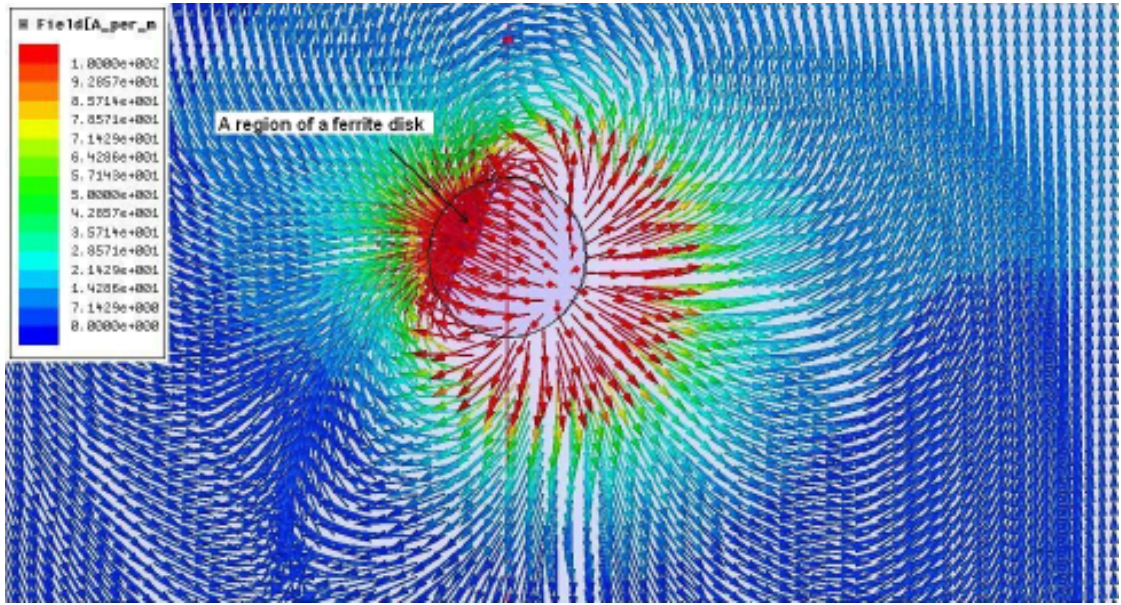

$\omega t = 0°$

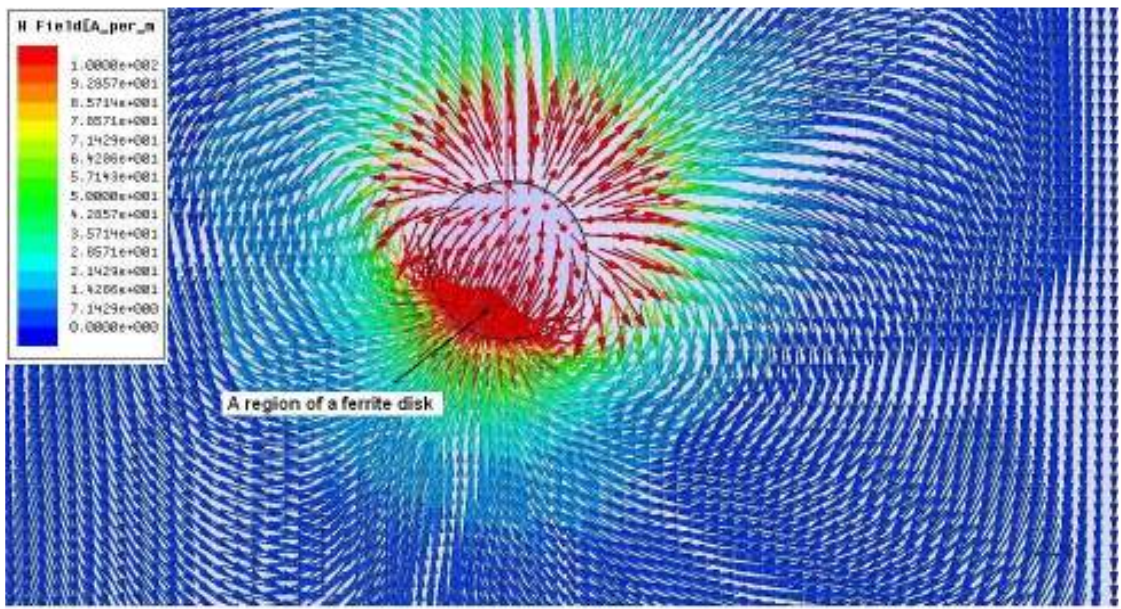

$\omega t = 90°$

Fig. 14. Magnetic field in a vacuum region (75 mkm above a ferrite disk) for the 1- resonance at different time phases.

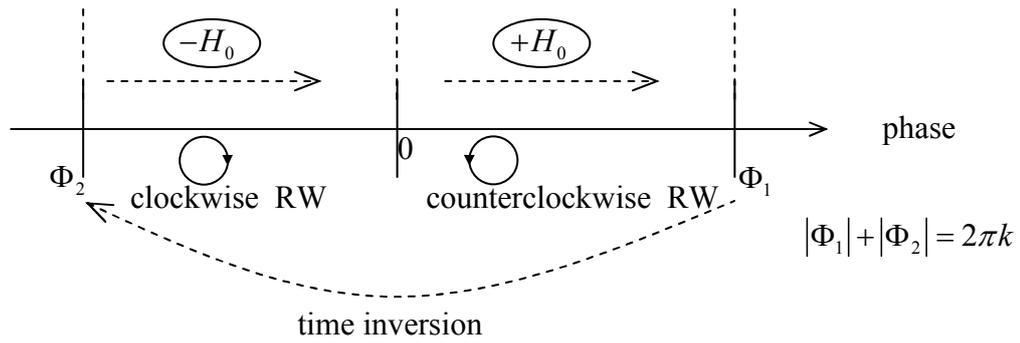

Fig. 15. The phases for counterclockwise and clockwise rotating waves (RW) in a MS-mode cylindrical resonator.

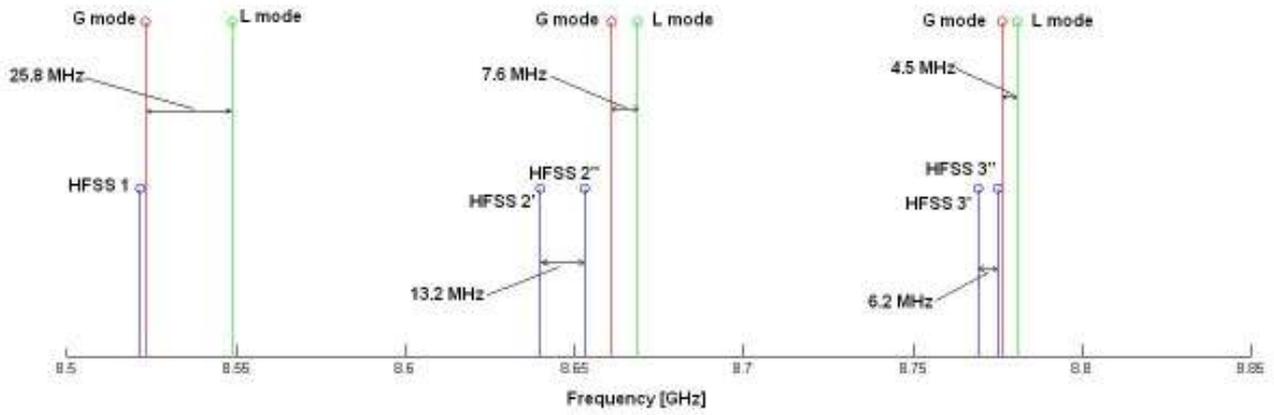

Fig. 16. The spectrum peak positions for the HFSS simulation and analytical *L*- and *G*-modes. Frequency differences for peak positions of the analytically derived *L* and *G* modes are at the same order of magnitudes as frequency differences for split-resonance states observed in the HFSS spectrum.